\documentclass[fleqn,usenatbib]{mnras}

\usepackage{newtxtext,newtxmath}
\usepackage[T1]{fontenc}

\DeclareRobustCommand{\VAN}[3]{#2}
\let\VANthebibliography\thebibliography
\def\thebibliography{\DeclareRobustCommand{\VAN}[3]{##3}\VANthebibliography}

\usepackage{graphicx}	% Including figure files
\usepackage{amsmath}	% Advanced maths commands
\title[Tides in the TRAPPIST-1 system]{Long-term tidal evolution of the TRAPPIST-1 system}

\author[R. Brasser et al.]{
R. Brasser,$^{1}$\thanks{E-mail: ramon.brasser@csfk.org}
G. Pichierri,$^{2}$
V. Dobos$^{3,4}$
and A.~C. Barr$^{5}$
\\
$^{1}$Origins Research Institute, Research Centre for Astronomy and Earth Sciences, Konkoly Thege Miklos St. 15-17, H-1121 Budapest, Hungary\\ MTA Centre of Excellence \\
$^{2}$Max Planck Institute for Astronomy, K\"{o}ningstuhl 17, D-69117 Heidelberg, Germany\\
$^{3}$Kapteyn Astronomical Institute, University of Groningen, Landleven 12, NL-9747 AD Groningen, Netherlands\\
$^{4}$MTA-ELTE Exoplanet Research Group, 9700, Szent Imre h. u. 112, Szombathely, Hungary\\
$^{5}$ Planetary Science Institute, 1700 East Fort Lowell, Suite 106, 85719 Tucson AZ, United States
}

\pubyear{2021}

\begin{document}
\label{firstpage}
\pagerange{\pageref{firstpage}--\pageref{lastpage}}
\maketitle

\begin{abstract}
The ultracool M-dwarf star TRAPPIST-1 is surrounded by seven planets configured in a resonant chain. Transit-timing variations have shown that the planets are caught in multiple three-body resonances and that their orbits are slightly eccentric, probably caused by resonant forcing. The current values of the eccentricities could be a remnant from their formation. Here we run numerical simulations using fictitious forces of trapping the fully-grown planets in resonances as they migrated in the gas disc, followed by numerical simulations detailing their tidal evolution. For a reduced disc scale height $h\sim 0.03$--0.05, the eccentricities of the planets upon capture in resonance are higher than their current values by factors of a few. We show that the current eccentricities and spacing of planets d to h are natural outcomes of coupled tidal evolution wherein the planets simultaneously damp their eccentricities and separate due to their resonant interaction. We further show that the planets evolve along a set of equilibrium curves in semimajor axis--eccentricity phase space that are defined by the resonances, and that conserve angular momentum. As such, the current 8:5--5:3--(3:2)$^2$--4:3--3:2 resonant configuration cannot be reproduced from a primordial (3:2)$^4$--4:3--3:2 resonant configuration from tidal dissipation in the planets alone. We use our simulations to constrain the long-term tidal parameters $k_2/Q$ for planets b to e, which are in the range $10^{-3}$ to $10^{-2}$, and show that these are mostly consistent with those obtained from interior modelling following reasonable assumptions.
\end{abstract}

\begin{keywords}
methods: numerical -- Planetary systems -- planets and satellites: terrestrial planets
\end{keywords}

%%%%%%%%%%%%%%%%%%%%%%%%%%%%%%%%%%%%%%%%%%%%%%%%%%

%%%%%%%%%%%%%%%%% BODY OF PAPER %%%%%%%%%%%%%%%%%%

\section{Introduction}
\begin{figure}
\resizebox{\hsize}{!}{\includegraphics[angle=0]{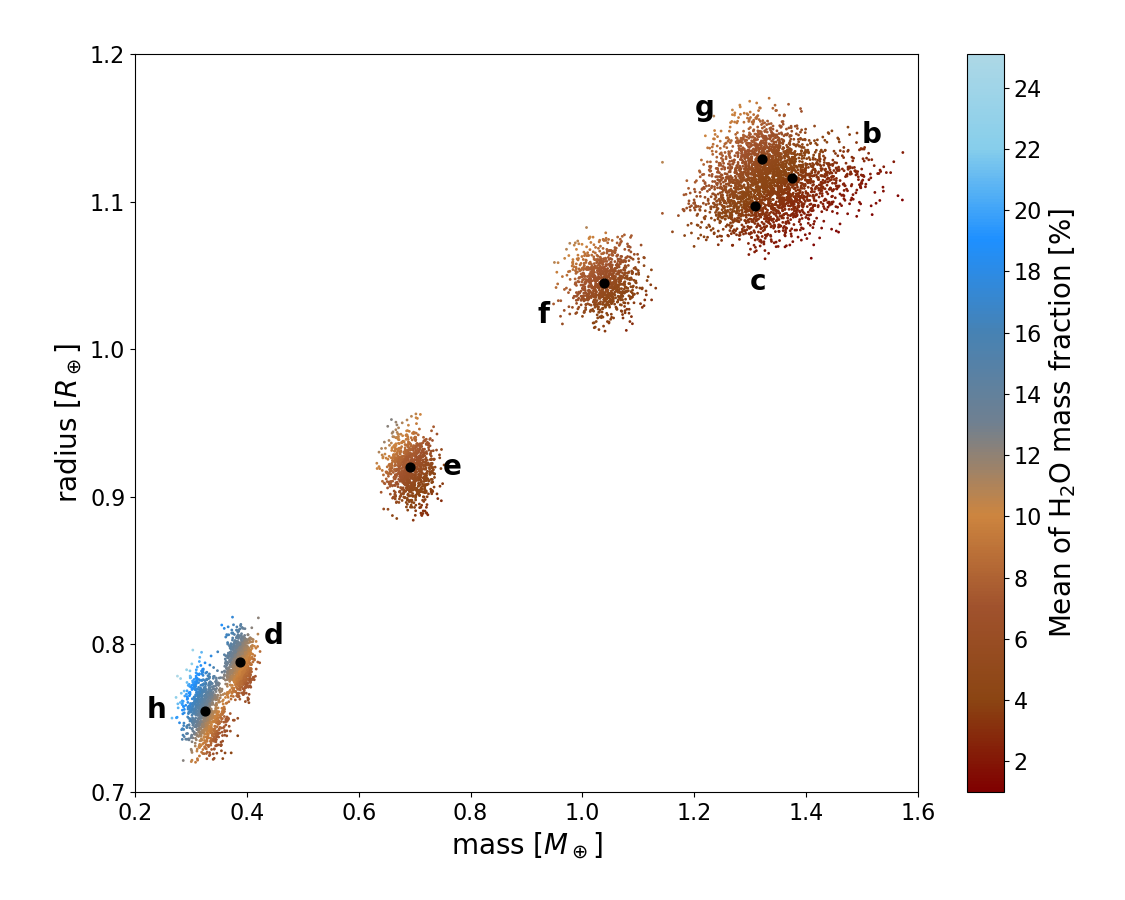}}
\caption{H$_2$O content (indicated by colours) for possible interior structures at a given mass--radius pair of the TRAPPIST-1 planets. The mean of the H$_2$O mass fraction is presented (including liquid water, ice I and high-pressure ice polymorphs) for all possible interior structures calculated for each mass--radius pairs. The calculation method used is the same as described in the work of \citet{Dobos2019} with the additional condition that only those interior structures were considered that contain 10$-$50~\% mass fraction of iron core. Masses and radii are from \citet{Agol2021} with black dots denoting the mean values.}
\label{fig:mrt1}
\end{figure}

The star TRAPPIST-1 is an ultracool M-dwarf that harbours seven roughly Earth-sized planets \citep{Gillon2017}. All of these planets orbit within 0.07 au of the star, and have orbital periods from 1.5 to $\sim$19 days \citep{Gillon2017,Grimm2018,Agol2021}. The planets are in a resonant chain, possibly involving all the planets, which results in the libration of a number of three-body resonant angles \citep{Luger2017}. The orbits of all planets appear to be mildly eccentric, with eccentricities $\lesssim 0.01$, although the inner two planets are consistent with having circular orbits \citep{Agol2021}. These eccentricities are possibly the result of forcing from the resonances or a remnant of their formation. Each of the planets has a density intermediate between the densities of compressed water ice and the Earth's inner core \citep{Gillon2017, Grimm2018,Barr2018,Agol2021}, implying solid planets composed of rock, metal \citep{elkins08}, and possibly ice if atmospheric conditions allow \citep{turbet2020} (see Fig.\ref{fig:mrt1}). With recent estimates for the planetary masses the system appears dynamically stable on Myr, and possibly Gyr, timescales \citep{Grimm2018,Agol2021}.

There is much uncertainty about the formation of the system, which is an active topic of research. Based on dynamical simulations \citet{Tamayo2017} suggested that the current resonant structure was the result of convergent migration within the primordial gas disc. That same year \citet{Ormel2017} suggested that the system could have been formed through a combination of planet migration and pebble accretion. More recently \citet{HuangOrmel2022} expanded on the pebble accretion idea with the addition of specific requirements for the interior structure of the disc and the configuration of the planets.

Short term ($\lesssim 1$~Myr) numerical simulations indicate that for many initial conditions within the observational uncertainties almost all of the resonant angles appear to librate \citep{Grimm2018,Brasser2019,Agol2021}. Their long-term ($>1$~Gyr) stability cannot be determined with the current observational uncertainties \citep{Agol2021}. From older observational data with larger uncertainties \citet{Brasser2019} found that most three-body resonances break on short timescales.

It is not clear whether the inner two planets, b and c, are truly involved in mean-motion resonances. Planets b and c are close to a mutual 8:5, and c and d in a 5:3 resonance \citep{Gillon2017,Luger2017}. This configuration warrants attention because the outer five planets all reside in first-order mean motion resonances; these two resonances are of second and third order, respectively. Indeed, \citet{Tey2022} showed that convergent migration in the gas disc leads to capture primarily in first-order mean-motion resonances, and that the 8:5 resonance between planets b and c, as well as the 5:3 between planets c and d, are difficult to explain via this mechanism. In this work we explore whether the current configuration can be obtained by tidal dissipation in the planets from an initial configuration consisting only of first-order mean-motion resonances because such a configuration is the expected outcome from planet migration. In the subsection below this is explained in more detail.

\subsection{Basic outline of resonant trapping}
\label{subsec:TrappingIntoResonance}
The formation of the TRAPPIST-1 planets is not yet well understood, but recent studies point to a potential combination of pebble accretion and planet migration \citep{Ormel2017,Unterborn2018,Schoonenberg2019} possibly followed by rebound because of the dispersion of the disc as well as tidal dissipation in the planets \citep{HuangOrmel2022}. An alternative scenario relies on different modes of migration \citep{Ogihara2022}. As the planets migrate through the disc towards the star they will encounter mutual mean-motion resonances \citep[e.g.][]{Petrovich2013}; this is especially so if the innermost planet stalls at the inner edge of the disc and the subsequent planet migrate towards it \citep[e.g.][]{Ogihara2010}.

The current configuration of the planets is in a suspected 8:5--5:3--(3:2)$^2$--4:3--3:2 resonant chain \citep{Gillon2017,Luger2017}, which can be obtained through planet migration \citep{Tamayo2017}. Yet other simulations of the formation of the system preferentially show trapping in a different configuration, e.g. (3:2)$^4$--4:3--3:2 \citep{Tey2022,HuangOrmel2022}. Here we briefly describe how resonant trapping works and why the latter configuration is more likely than the former.

Away from resonances, the averaged disturbing function that describes the orbital evolution of two planets is the secular part, which is given by \citep{murray99}

\begin{equation}
\mathcal{R}_D^{\rm sec} = {\textstyle \frac{1}{8}} \alpha b_{3/2}^{(1)}(\alpha)(e^2+e'^2)-{\textstyle \frac{1}{4}}\alpha b_{3/2}^{(1)}(\alpha)ee'\cos(\varpi-\varpi'),
\end{equation}
where $e$ and $e'$ are the eccentricities of the inner and outer planet, $\varpi$ and $\varpi'$ are their longitudes of periastron, $\alpha=a/a'$ is the ratio of semi-major axes, and $b_s^{(j)}$ are the so-called Laplace coefficients, which can be expressed in terms of Gamma- and hypergeometric  functions \citep{murray99}. In a two-or-more-planet system each planet has a free eccentricity and a forced eccentricity caused by their mutual perturbations. The motion of the eccentricities and longitudes of perihelia of a two-planet system are coupled via
\begin{eqnarray}
e_1 \exp(\imath\varpi_1) = M_{1,1}\exp[\imath(g_1t+\beta_1)] + M_{1,2}\exp[\imath(g_2t+\beta_2)], \nonumber \\
e_2 \exp(\imath\varpi_2) = M_{2,1}\exp[\imath(g_1t+\beta_1)] + M_{2,2}\exp[\imath(g_2t+\beta_2)].
\end{eqnarray}
The $M_{i,i}$ terms are the free eccentricities, while the $M_{i,j}$ terms are the forced eccentricities, and $t$ is the time. The frequencies $g_i$ are the eccentricity eigenfrquencies with their corresponding phases $\beta_i$, and $\imath^2=-1$ is the imaginary unit.

Near a mean-motion resonance the angles of the form $\phi=j_1\lambda + j_2 \lambda' + j_3 \varpi + j_4 \varpi'$, where $j_1,\dots,j_4$ are integers, will vary slowly, on timescales much longer than the orbital periods of the planets, and cannot be eliminated from the problem through averaging. The angle $\phi$ satisfies the D'Alembert rule that $\sum_i j_i = 0$.

When planets are in a mean-motion resonance the resonant disturbing function will typically be of the form \citep{murray99}

\begin{equation}
    \mathcal{R} = \mathcal{R}_D^{\rm sec} + e^{|j_3|}e'^{|j_4|}(f_d(\alpha)+f_e(\alpha)\cos \phi),
\end{equation}
where $f_d(\alpha)$ and $f_e(\alpha)$ are functions of Laplace coefficients. It is immediately obvious that the strength of the resonant perturbation depends on the product $e^{|j_3|}e'^{|j_4|}$ i.e. on $j_3+j_4$, which itself depends on $j_1+j_2$. For a first-order resonance, such as the 2:1, the 3:2 or the 4:3, we have $|j_1+j_2|=1$ and thus $|j_3+j_4|=1$, and $\mathcal{R} \propto e$ or $\mathcal{R} \propto e'$. For a second-order resonance, such as the 3:1 or the 5:3 we have $|j_1+j_2|=2$, so that by necessity $|j_3+j_4|=2$ and either $\mathcal{R} \propto e^2$, or $\mathcal{R} \propto ee'$, or $\mathcal{R} \propto e'^2$. For a third-order resonance, such as the 4:1 or the 8:5, $\mathcal{R}$ is proportional to the third power in the eccentricities, and so forth. A resonant pair of planets will have non-zero forced eccentricities if they are in equilibrium.

The interaction of the planets with the gas disc causes them to migrate towards the star and to have their free eccentricities damped \citep{Tanaka2004} i.e. the $M_{i,i}$ terms will approach zero - and by necessity so will the $M_{i,j}$ terms because they depend indirectly on the $M_{i,i}$ terms. In essence the planets approach the mean-motion resonance with (close to) zero free eccentricities. For low eccentricities the perturbations from first-order resonances will far exceed those from second- or third-order resonances due to the argument $e^{|j_3|}e'^{|j_4|}$ in the disturbing function. Capture in (first-order) resonance is facilitated by lower eccentricities \citep{Henrard1982}, so that the {\it expectation} is that the planets will be trapped into {\it first-order} mean-motion resonances when they are migrating starwards, rather than into higher-order resonances.

Once the planets are captured into first-order mean-motion resonances while they are migrating, each pair of planets will approach the resonance with a period ratio greater than the exact resonant value, and their eccentricities will increase due to the increasing importance of the resonant perturbations over the secular perturbations. The resulting eccentricities are forced eccentricities. The planets will reach an equilibrium between the eccentricity pumping from the resonance and the damping of the gas disc, with the equilibrium eccentricity being approximately $e \sim 1.3h$ \citep{Goldreich2014,2018CeMDA.130...54P,Brasser2019} where $h$ is the reduced scale height of the gas disc. 

What we currently observe is that the outer five planets are in a 3:2--3:2--4:3--3:2 chain \citep{Luger2017} with eccentricities below 0.01, but the inner three are in a potential 8:5--5:3 chain \citep{Luger2017}, which is of third and second-order respectively. The probability of capturing the planets in these higher-order resonances during the migration phase is very low \citep{Papaloizou2018}. \citet{Tamayo2017} were able to migrate all seven planets into the current resonant chain by starting the planets a few percent wide of their current resonances and only applying migration torques to planet h. In contrast, \citet{HuangOrmel2022} advocate a primordial (3:2)$^4$--4:3--3:2 configuration that was disrupted by rebound from the dispersing gas disc followed by tidal dissipation in the planets. In the next section, we introduce a resonant model for such a chain which encapsulates the main dynamical effects of planetary migration and planetary tides on the planets.

\subsection{Resonant loci for a TRAPPIST-1 (3:2)$^4$ -- 4:3 -- 3:2 chain}
\begin{figure}
\resizebox{\hsize}{!}{\includegraphics[angle=0]{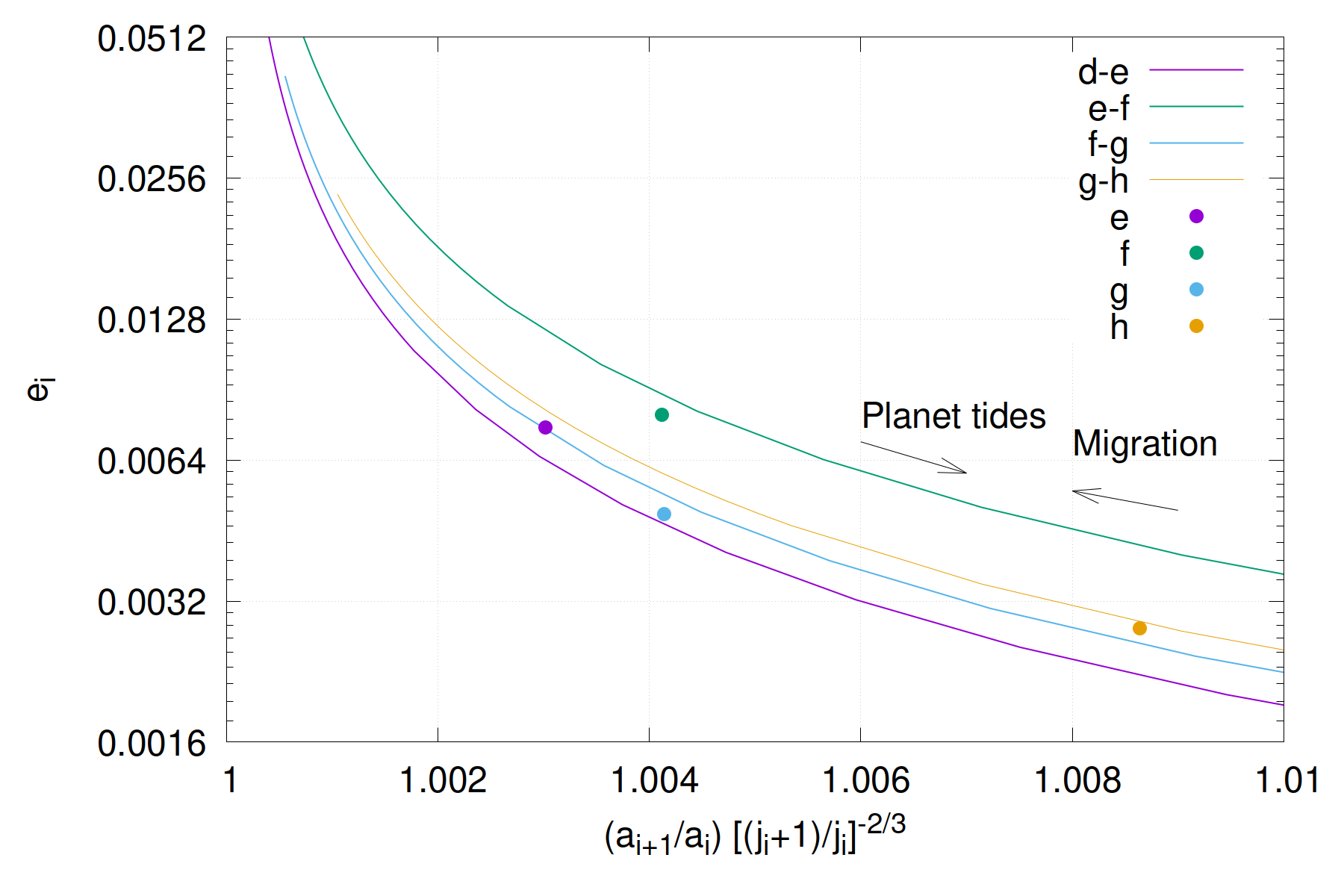}}
\caption{Normalised resonant loci of several resonances in the TRAPPIST-1 system. The dots show the current configuration of the system. Arrows indicate the direction of motion along the lines for disc-induced migration, and for tidal dissipation in the planets.}
\label{fig:eq}
\end{figure}

The evolution of the system is readily derived from the parametrisation shown in Fig.~\ref{fig:eq}. All planet pairs will evolve along equilibrium lines that are determined by the Hamiltonian of the system. The resonant Hamiltonian model for any resonance or resonant chain can be constructed following the recipe presented in \citet{2013A&A...556A..28B} for two planets, and extended to the multiple-planet case. For example, the framework presented in \citealt{2019A&A...625A...7P} applies to three resonant planets; this procedure is extended here to encompass the 7-planet case of TRAPPIST-1. Even when the Hamiltonian is expanded to first order in the eccentricities -- which is justified by the low eccentricities of the TRAPPIST-1 planets -- the resulting Hamiltonian is not integrable when there are more than two planets, and the resonant equilibria cannot be found analytically. As such, one has to resort to numerical optimisation methods to obtain the location of the equilibria of the resonant system, which we call resonant equilibrium points. We implemented the resulting semi-analytical scheme in the \texttt{Mathematica} language and used the \texttt{FindRoot} function to find the extrema of the Hamiltonian\footnote{
We note that we included only interactions between neighbouring planets in the resonant model. Given the large number of variables (there are $2(N_\mathrm{pl}-1))$ degrees of freedom in a resonant chain of $N_\mathrm{pl}$ planets, therefore $4(N_\mathrm{pl}-1)=24$ variables for the full TRAPPIST-1 chain), this approach becomes numerically more feasible since the stable equilibria we are interested in only occur for values of the resonant arguments equal to 0 or $\pi$, thus cutting by half the number of equations to be solved numerically. This approach does not consider the 2:1 contribution between planets e and g, which would shift the loci only slightly in the semi-major-axis ratio v.s. $e$ space, without however changing the main feature of the resonant curves used in this paper}.

Since the problem is scale free, let us think in terms of semi-major-axis ratios between the planets; this is equivalent to having fixed units of length such that the nominal resonant location of planet b is e.g.\ $a_\mathrm{b,res} = a_\mathrm{b,obs}$, i.e. its currently observed value. For a fixed value of the total angular momentum of the system one can find one stable resonant locus, which then determines the semi-major axes and eccentricities of all planets relative to each other via the equilibrium resonant condition. By varying the value of the total angular momentum one gets a whole family of equilibrium points, parametrised by the angular momentum of the system (or, equivalently, by any one of the eccentricities or semi-major axes of a specific TRAPPIST-1 planet). 

Such scale-free equilibrium curves are displayed in Fig.~\ref{fig:eq} on a plane with $(a_{i+1}/a_i) \times [(j_i+1)/j_i]^{-2/3}$ on the horizontal axis vs.\ the eccentricity $e_i$ on the vertical axis ($i$ being a label for each pair of subsequent planets along the chain), for the four outermost planets. Here, $j_i$ are the indices of the $(j_i+1):j_i$ mean-motion resonances along the (3:2)$^4$ -- 4:3 -- 3:2 chain, namely $j_1=j_2=j_3=j_4=j_6=2$ and $j_5=3$. The main feature to note of these curves is that at decreasing eccentricities the resonant loci deviate away from the exact commensurability $(a_{i+1}/a_i) \times [(j_i+1)/j_i]^{-2/3}=1$; this effect is well known (e.g.\ \citealt{Delisle2012,2019A&A...625A...7P}) and is due to the faster precession of the pericentres at vanishing $e$ because to first order $\dot{\varpi} \propto e^{-1}$. The resonant locus for a single planet pair is linked to corresponding loci for all other planet pairs by the total angular momentum of the system. Therefore, in a resonant chain, lowering the eccentricity of one planet along the curve forces all other eccentricities to decrease as well. The dots show the current normalised semi-major axis ratio and eccentricities of planets e to h. Migration induced by disc torques will cause each planet to move leftwards along the equilibrium lines while planetary tides will cause them to move in the opposite direction \citep{Delisle2012,2019A&A...625A...7P}.

Here we investigate the tidal evolution of the system starting from both the current resonant chain and a (3:2)$^4$--4:3--3:2 configuration with the aim to constrain the tidal parameters in some of the inner four planets. We further aim to establish whether the current state can be reproduced through tidal dissipation in the planets from a primordial (3:2)$^4$--4:3--3:2 configuration. This problem was partially studied by \citet{Papaloizou2018} but they used old values for the planetary masses and radii. Here we make use of the latest masses and radii from \citet{Agol2021}.

\section{Methodology}
\subsection{Resonant trapping}\label{subsec:ResonantTrapping}
The first step in our analysis of possible past and present resonant states for the TRAPPIST-1 requires the build-up of a resonant chain. \citet{Tamayo2017} showed that a TRAPPIST-1 system with period ratios close to the current observed ones can be obtained by implementing inward migration applied only to the outermost planet TRAPPIST-1h, thus hinting to a dynamical past marked by planet-disc interactions reminiscent of type-I migration. We should keep in mind that this approach to explain the precise formation history of the system is probably too simplistic, since migration should be applied to all planets at once, which, given their non-uniformity in planetary mass along the chain, would actually result in both convergent and divergent migration for specific planet pairs. Indeed, more sophisticated formation histories seem to be necessary to form an initial resonant state close to the current one \citep[e.g.][]{HuangOrmel2022}. For the purposes of our analysis, however, we make use of a recipe similar to that proposed by \citet{Tamayo2017} only so that we can reliably form a desired chain, in order to then study its long-term tidal evolution after gas disc dissipation. Indeed, the advantage of the approach of \citet{Tamayo2017} is that migration only involves planets that have already assumed a resonant relationship with planets farther out, which makes the process more controllable. It also allows the innermost planets to be captured in the high-order 8:5 and 5:3 resonances, even if one would actually expect first-order resonances to be favoured as explained in the Introduction. This trapping behaviour can be observed in the work of \citet[e.g.][]{Tey2022} and is confirmed by our own simulations of the innermost 3- and 4-planet system under conventional type-I migration prescriptions obtained from \citet{2008A&A...482..677C}. The trapping in the higher-order resonances is facilitated because the planets approach these higher order resonances with already non-zero eccentricities induced by their resonant interactions with planets farther out.

For our migration simulations we used the N-body integrator SyMBA \citep{Duncan1998} modified to include the forces from a gas disc and migration, as detailed below. We consider type-I inward migration applied to planet h only, with a migration timescale taken from \citet{2008A&A...482..677C} in the planar case:
\begin{equation}
\tau_\mathrm{mig} = 2 \frac{\tau_\mathrm{wave}}{(2.7 + 1.1 \alpha_\Sigma)} h^{-2}\left(P(e) + \frac{P(e)}{|P(e)|} \right),
\end{equation}
where
\begin{equation}
P(e) = \frac{1+\left(\frac{e}{2.25 h}\right)^{1/2} + \left(\frac{e}{2.84 h}\right)^6}{1-\left(\frac{e}{2.02 h}\right)^{4}}.
\end{equation}
We instead apply eccentricity damping to all planets, following the formula from \citet{2008A&A...482..677C} in the planar case:
\begin{equation}
    \tau_e = \frac{\tau_\mathrm{wave}}{0.780}\left[1-0.14\left(\frac{e}{h}\right)^2+0.06\left(\frac{e}{h}\right)^3\right].
\end{equation}
In these formul\ae\ $h=H/r$ is the aspect ratio of the disc, and $\tau_\mathrm{wave}$ is the wave timescale given by \citep{Tanaka2004}
\begin{equation}
    \tau_\mathrm{wave} = \left(\frac{M_*}{m_\mathrm{pl}}\right) \left(\frac{M_*}{\Sigma_\mathrm{gas,pl} a_\mathrm{pl}^2}\right) h_\mathrm{pl}^4 \Omega_{K,\mathrm{pl}}^{-1},
\end{equation}
where $m_\mathrm{pl}$ is the mass of the planet, $M_*$ that of the star, and $\Omega_{K,\mathrm{pl}}$ is the frequency of Keplerian motion, and all quantities with the label ``pl'' are evaluated at the location of the planet. 
We used aspect ratios values of $h=0.04$ and $h=0.05$, which has the effect of obtaining slightly different capture eccentricities, since $e_\mathrm{capt.}\approx \sqrt{\tau_e/\tau_a}\approx 1.3 h$ (\citealt{Goldreich2014,2018CeMDA.130...54P}, see also Subsect.\ \ref{subsec:TrappingIntoResonance}). The same formula shows that $\tau_\mathrm{wave}$ does not influence the final captured state, and thus neither does the surface density $\Sigma_\mathrm{gas,pl}$, provided that migration is slow enough to allow for adiabatic resonant capture. The masses of the individual planets used in this simulation are 1.43,
1.34, 0.37, 0.68, 1.02, 1.29 and 0.31~$M_\oplus$.

\subsection{SWIFT MVS FMA}
After trapping the planets into a resonant chain we need to establish how close the planets are to actual resonances. It is not sufficient to look at the evolution of the semi-major axes or the resonant angles, because the latter can still librate far from resonance if the system is tidally evolved \citep{Delisle2012,BatyginMorbidelli2013}. In most N-body codes, such as SWIFT MVS \citep{LD1994} or {\it Mercury} \citep{Chambers1999}, the semi-major axis output are osculating values that depend on the relative positions of the planets to each other. These osculating values will not yield high precision orbital frequencies without filtering. As such, we have taken a different approach and implemented direct Fourier analysis in the N-body symplectic integrator SWIFT MVS \citep{LD1994} as follows. 

During the simulations the $x$ and $y$ co-ordinates of the planets are stored in arrays, which are thinned by a factor of four to conserve memory. The MVS method relies on the symplectic mapping of \citet{WH1991}, which requires at least twenty steps per orbit to preserve accuracy. Our thinning therefore leaves at least five coordinate values for the orbit of the innermost planet which, through experimentation, we have determined to be enough to ensure an accurate computation of the orbital frequency. The size of the arrays is an input parameter, which needs to be equal to a power of two, and which has a current maximum dimension of 262,144 (although this can be increased if needed). Once the arrays are full, their data are passed to a subroutine which uses the frequency-modified Fourier transform (FMFT) method of \citet{SN1996} to compute the orbital frequencies of the planets with the corresponding amplitudes -- which for low-eccentricity orbits are approximately equal to the semi-major axes -- and initial phases. The FMFT routine by 
\citet{SN1996} contains a Hanning window filter that removes short period oscillations and cleans the signal. The output is written to a file as the simulation runs.

Once the FMFT subroutine has finished the simulation carries on until the arrays are full again, the frequencies are computed again over the next interval and written to disk, and the process is repeated until the simulation ends. At the end of the simulation, there exists a file whose output contains the orbital frequencies of all the planets computed over a series of time intervals of at most 1,048,576 time steps for each interval. One may then apply Laskar's Frequency Map Analysis \citep{Laskar1993} to the data.

\begin{figure*}
\resizebox{\hsize}{!}{\includegraphics[angle=0]{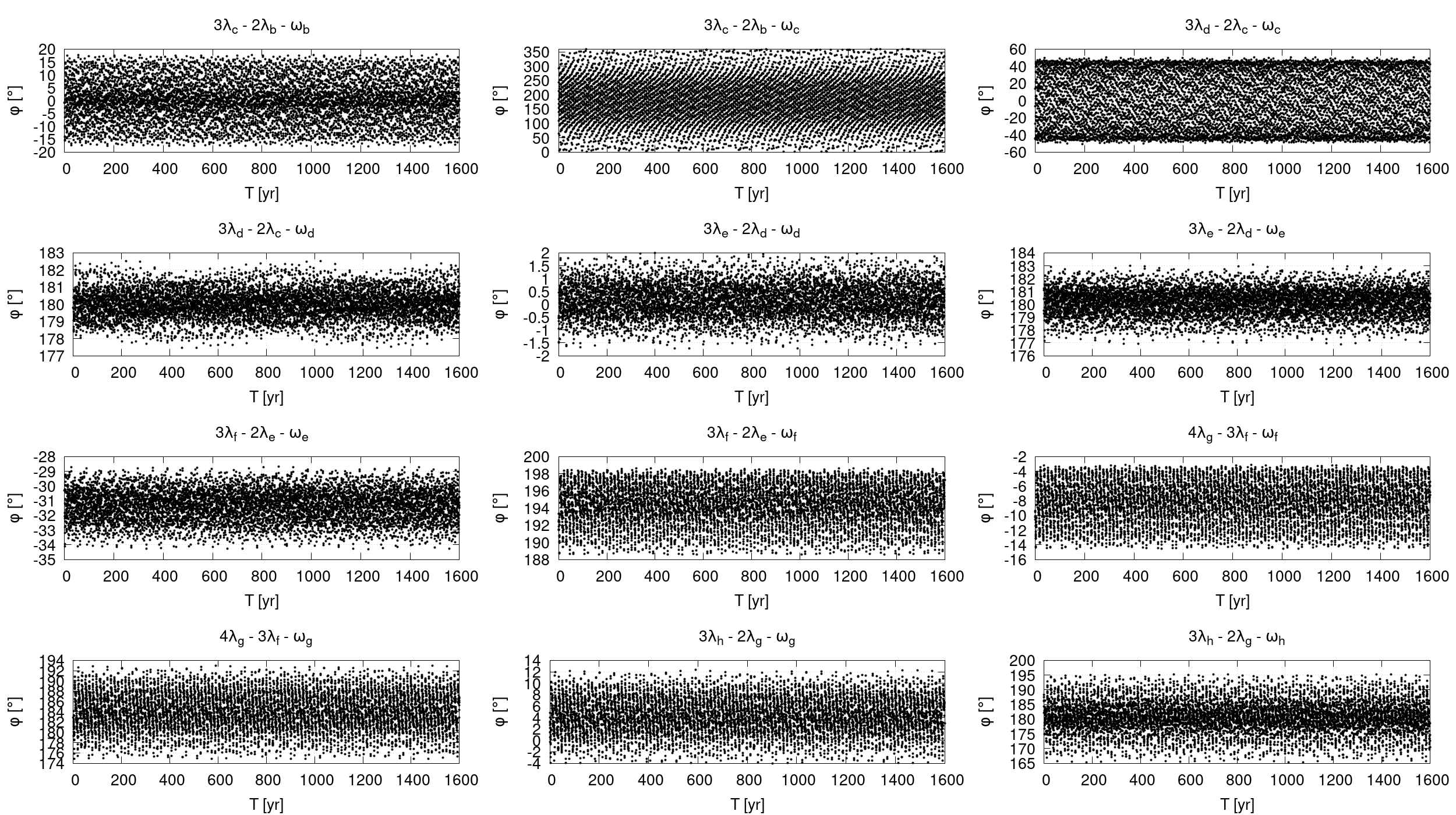}}
\caption{Evolution of resonant angles after migration through the gas phase into a (3:2)$^4$--4:3--3:2 chain. Note the different scales on the vertical axes associated to each angle. All planets are in resonance with each other, but the argument $3\lambda_c - 2\lambda_b - \varpi_c$ circulates. For this configuration $\nu = -2.64006$ rad/yr apart from the b-c pair for which $\nu = -29.8$ rad/yr.}
\label{fig:resarg}
\end{figure*}

\begin{figure*}
\resizebox{\hsize}{!}{\includegraphics[angle=0]{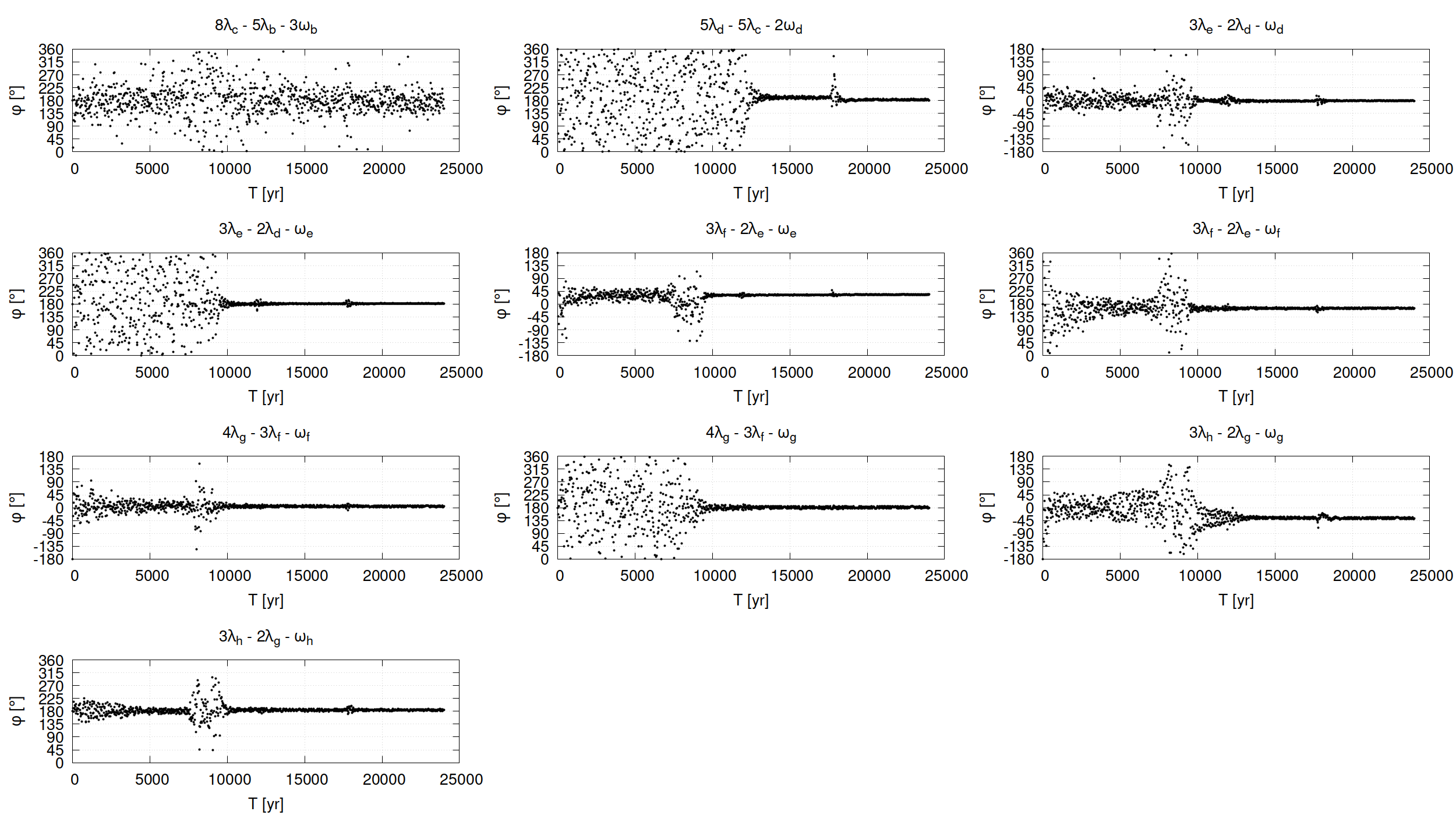}}
\caption{Resonant angle evolution during migration while attempting to reproduce the current configuration, close to a 8:5--5:3--3:2--3:2--4:3--3:2 chain. Each panel shows the evolution of a resonant angle associated with pairs of planets, for period ratios close to the observed ones. All angles are seen to be librating at the end of the simulation.}
\label{fig:resargmig}
\end{figure*}

\subsection{Application of SWIFT MVS FMA: analysis of the current system}
We have simulated 1250 initial conditions of the TRAPPIST-1 system from \citet{Agol2021}. For the analysis presented below we picked one system wherein the resonant angles of the outer five planets librated with low amplitude, because this is the expected outcome from a system that migrated into resonance in the presence of a gas disc \citep{Delisle2012}. A resonant chain is characterised by all the mean motions of the planets being near resonance, such that $\nu = (j+1)n' - jn$ is a relatively slowly varying variable for each pair of planets, with a period much longer than the individual orbital periods of the planets. We find that for all the five outer resonances $\nu=-4.68006$ rad/yr. To maintain the resonance implies that the longitudes of periastrion of the five outer planets all regress with the same frequency i.e. for these planets we expect that $\dot{\varpi}=-4.68006$ rad/yr. A Fourier analysis of the vectors $e\exp(I\varpi)$ indicates that this is the case: the longitudes of periastron of planets d to h indeed all regress with the same frequency $\nu$. 

However, libration of the resonant angles does not necessarily imply that the planets are in resonance. According to \citet{Delisle2012} the libration could be a geometric projection effect caused by tidal separation. One way to check this is to decompose the quantity $e\exp(\imath\phi)$, where $\phi=(j+1) \lambda' -j\lambda - \varpi$ or $\phi=(j+1) \lambda' -j\lambda - \varpi'$. \citet{Delisle2012} claim that in a truly resonant case $(j + 1)n' - jn \approx 0$, and thus $(j + 1)\lambda' - j\lambda$ has a long period, and the $\varpi$ are dominated by the secular eigenmodes, which also have long periods. If the libration of the angles is a projection effect due to tidal separation, then $(j + 1)\lambda' - j\lambda$ and $\varpi$ are dominated by short periods (high frequencies), with frequencies equal to $\nu - g$. In Table~\ref{tab:freqs} we show frequency output for planet d from a set of initial conditions wherein all the resonant angles of the outer five resonances librated (see Fig.~\ref{fig:resarg}). It is clear from this Fourier decomposition that the current eccentricity of planet d is a purely forced eccentricity caused by its proximity to resonance with planet e. We find that the dominant terms in the decomposition of $e_d\exp[\imath(3\lambda_e-2\lambda_d-\varpi_d)]$ have frequencies $-4.792$ rad/yr and $-4.860$ rad/yr, which are equal to $\nu -g_3$ and $\nu-g_4$ respectively. For this argument the term with the largest amplitude, however, has a frequency close to zero because it is caused by the forced eccentricity due to the resonance \citep{Delisle2012}. Similar results are obtained for the other resonant angles. Thus, for this specific set of initial conditions we used here, the planetary configuration appears to be resonant, but it is likely to be a geometric projection. This is the expected outcome of migration into a resonance with near-zero free eccentricity \citep{Delisle2012}. We have not applied such a rigorous analysis to other initial conditions from \citet{Agol2021} in part because the resonant angles do not always librate. However, we expect that from analysing these systems the conclusions will be similar.

\begin{table}
	\centering
	\caption{Frequency decomposition of planet d. The leftmost columns shows the frequencies of $e_d\exp(\imath \varpi_d)$ and the rightmost columns those of $e_d\exp[\imath(3\lambda_e-2\lambda_d-\varpi_d)]$. The amplitude columns are essentially the values of the forced eccentricity of planet d.}
	\label{tab:freqs}
	\begin{tabular}{lc|cc} % four columns, alignment for each
	freq [rad/yr] & amplitude [$10^{-3}$] & freq [rad/yr] & amplitude [$10^{-3}$]\\ \hline \\
-4.68006 & 4.10185 &       -6.5363$\times 10^{-7}$ & 4.11497  \\
0.112218 ($g_3$)& 0.477816  & -4.79227 ($\nu-g_3$) & 0.482074 \\
0.180291 ($g_4$) & 0.389516  & -4.86035 ($\nu-g_4$) & 0.397321 \\
0.168342 ($g_2$) & 0.230608 & -4.84841 ($\nu-g_2$) & 0.236361 \\
\end{tabular}
\end{table}

\subsection{Tidal evolution}
In order to investigate the long-term effect of tidal damping in the planets on their orbits, we have evolved the TRAPPIST-1 planetary system subjected to planetary tides with the SyMBAt code \citep{PB2018}, which is based on Mercury-T \citep{Bolmont2015}. We analyse the tidal evolution of two distinct sets of initial conditions and apply dissipation in one or multiple planets within these initial conditions. The initial conditions consisted of both the current system and the (3:2)$^4$--4:3--3:2 chain; the latter initial conditions were obtained with migration in the gas disc. The parameter space for tidal dissipation in the planets is large due to the high number of planets and the potentially large range in $k_2/Q$ per planet, where $k_2$ is the Love number and $Q$ is the tidal quality factor. To keep the number of simulations reasonable we have taken a simplified approach. To keep the running time of the simulations reasonable we have fixed the value of $Q=10^{-3}$ for each planet, while the $k_2$ values were taken from a best-fit between the values of Earth and Mars as a function of planetary mass, which results in $k_2=0.29(m_p/M_\oplus)^{0.26}$. For each set of initial conditions we ran 7 simulations. In the first four the damping occurred only in a single planet b to e. In the fifth to seventh simulations damping occurred in planets b and c, b to d and b to e respectively. Each simulation was run for 1 Myr with a time step of $10^{-4}$~yr. A second set of simulations was run with $Q=10^{-2}$ for 10 Myr to verify whether the results with $Q=10^{-3}$ were not subjected to excessive damping.

\subsection{Internal tidal models}
To determine the possible interior structures of the planets, the model described by \citet{Dobos2019} was used. This model assumes an iron core, a rock mantle, a high-pressure ice polymorph layer and either liquid water or ice I state on the surface. We set the iron core to have a mass fraction between 20 and 40~\% of the total mass of the body, to achieve a realistic restriction on probable interior structures.

The mass, radius, and semi-major axis of the orbit are adapted from \citet{Agol2021}. Their uncertainties are used in a Monte-Carlo approach to generate a thousand sets of mass--radius--semi-major axis values assuming a Gaussian distribution around the mean value. For each set of parameters several interior structure models are generated, and on average we compute 16,000--25,000 solutions for each planet (a growing number from planet b to f) for a given orbital eccentricity. This provides a large range of possible solutions that preserve the most likely outcomes for the interior compositions.

Using the model of \citet{Dobos2019}, the tidal heating (generated by the tidal forces from the star) is calculated for each interior structure generated. This is a viscoelastic tidal heating model with a Maxwell rheology, mimicking the effects of each layer. We assume a temperature-dependent rheology that includes the effect of melt on viscosity and shear modulus \citep{henning09, Barr2018, Dobos2019}. The $k_2/Q$ tidal parameter is then calculated following the description of \citet{Brasser2019}.

\section{Results}
\subsection{Migration in the gas disc}
We implemented the disc-planet interaction prescriptions described in Subsection \ref{subsec:ResonantTrapping} to a 7-planet TRAPPIST-1 system. We use masses drawn from \cite{Agol2021}, thus assuming that the planets have already fully formed and interact with the gaseous disc. We start each simulation with all planet pairs at period ratios 2\% wider than either their nominal resonant values for their currently observed 8:5--5:3--(3:2)$^2$--4:3--3:2 configuration, or the nominal resonant (3:2)$^4$--4:3--3:2 configuration. In all cases, planets' orbits are initiated on a co-planar, circular configuration. Following the recipe of \cite{Tamayo2017} planet h is the only one feeling a negative torque from the disc, and it migrates inward. It is captured into the desired 3:2 resonance with planet g, and the two migrate in resonance inward towards planet f. Planet g is subsequently captured in the desired 4:3 resonance with planet f, and the process continues. This numerical setup allows us to form both an 8:5--5:3--(3:2)$^2$--4:3--3:2 chain and a (3:2)$^4$--4:3--3:2 chain for the TRAPPIST-1 system. The evolution of the resonant angle for each resonance during the migration attempting to reproduce the current resonant chain is shown in Fig.~\ref{fig:resargmig}. The (3:2)$^4$--4:3--3:2 chain has been suggested by \citet{HuangOrmel2022} to be a natural outcome of type-I migration. Furthermore \citet{Tey2022} show that trapping in multiple 3:2 resonances is the expected outcome, even for the b-c and c-d pairs. 

When we formed the (3:2)$^4$--4:3--3:2 chain we found that planet b was not always deeply in resonance. In order to obtain a system deeper into the resonance for planet b, we also apply artificial migration to it, but smoothly reverse the sign of the torque from negative to positive around a location close to the current observed location of the planet. In this manner we simulate a simple planet trap (\citealt{2006ApJ...642..478M}, using the prescription from \citealt{2018CeMDA.130...54P}).
%\begin{figure*}
%\resizebox{\hsize}{!}{\includegraphics[angle=0]{eccSet2.png}}
%\caption{Tidal evolution Set 2.}
%\label{fig:ecc1}
%\end{figure*}

\subsection{State of the system after migration in the gas disc}
The first set of initial conditions that arose from the gas-driven migration simulations attempted to reproduce the current configuration, i.e. the 8:5--5:3--(3:2)$^2$--4:3--3:2 chain. We confirm that the resonant angles of the last four resonances are librating, but not those of the inner two pairs. The frequency for the outer four resonances is found to be $\nu = -1.74400$ rad/yr, which is lower than the current value of $\nu = -4.68007$ rad/yr. This is expected because the closer the planets migrate to the exact resonance the closer the frequency $\vert \nu \vert$ will be to zero. After migration the planets were closer to exact resonance so that their forced eccentricities are also higher than the current values (see Fig.~\ref{fig:eq} for how the forced eccentricity changes as a function of distance to exact resonance) and their present eccentricities are probably the result of tidal damping in the planets.

We created a second set of initial conditions where the planets were migrated into resonance in the presence of a gas disc that consists of a (3:2)$^4$--4:3--3:2 chain, because other migration experiments show that this is a likely outcome \citep{HuangOrmel2022,Tey2022}. We have plotted all of the resonant angles of this chain in Fig.~\ref{fig:resarg} and it is clear that apart from $3\lambda_c - 2\lambda_b - \varpi_c$ all of these angles are librating. For this particular resonant chain we computed $\nu=-2.64006$ rad/yr. The longitudes of periastron of planets c to e all regress with the same frequency i.e. for these planets $\dot{\varpi} = -2.64006$ rad/yr. An exception is planet b, for which $\dot{\varpi} = 1.607$ rad/yr and $\nu_{\rm bc} = -29.809$ rad/yr. It appears as if the innermost pair of planets is not truly part of the resonant chain, and all our attempts at trying to do so have failed.

\subsection{Tidal evolution of the current system}
\begin{figure*}
\resizebox{\hsize}{!}{\includegraphics[angle=0]{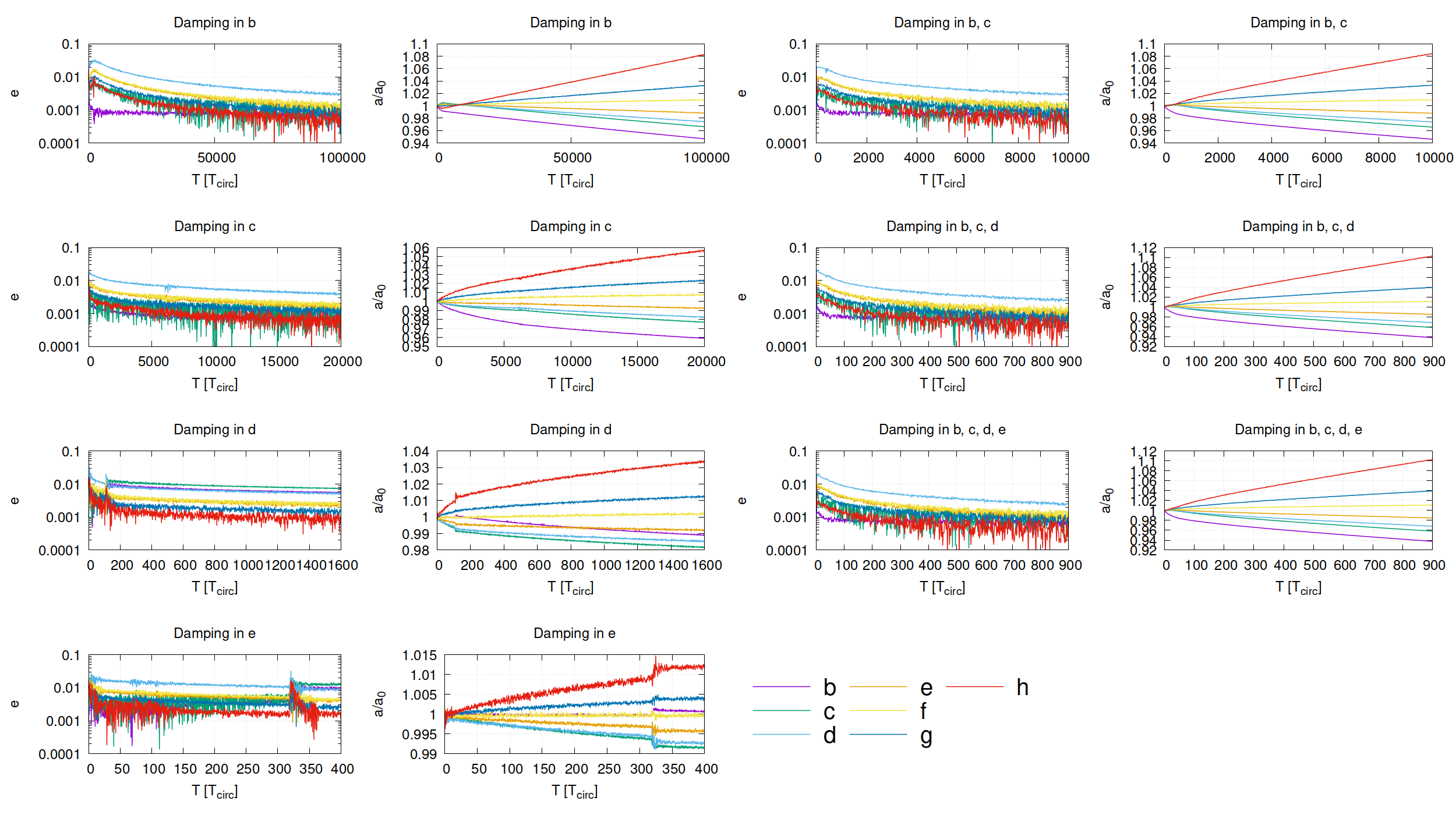}}
\caption{Tidal evolution of the current orbital configuration. The first and third columns show the eccentricities versus time while the second and fourth columns are the ratio $a/a_0$, where $a_0$ is the initial semi-major axis. Colours are purple for planet b, green for c, light blue for d, orange for e, yellow for f, deep blue for g and red for h.}
\label{fig:ecc4}
\end{figure*}
We have simulated the evolution of the resonant chain after gas-driven migration with planetary tides included using SyMBAt. The outcome of several attempts is shown in Fig.~\ref{fig:ecc4}. The time axis is scaled to the circularisation time of the planet where the damping occurs; for damping in b and c we chose the circularisation time for planet c and for damping in b, c, d and b, c, d and e we chose the circularisation time for planet d. 

We show the evolution of the eccentricities (first and third columns) and the ratio $a/a_0$ with time (second and fourth columns) for various configurations of damping; here $a_0$ is the semi-major axis at the beginning of the simulation. There are two immediate visible trends.

First, the eccentricities of planets d to h damp due to dissipation in (some of) the (other) planet(s). Second, all planets separate from each other due to angular momentum conservation. The inner four planets migrate inwards while the outer three migrate outwards. The separation is clearly visible because the distance between the lines increases as time increases. Since planets b and c are only weakly tied to the resonant chain consisting of planets d to h we see that the eccentricities of planets b and c stay low, while that of the outer five planets declines. 

An alternative manner to display the tidal evolution is to make use of the equilibrium curves, as is done in Fig.~\ref{fig:eq}, where we have computed the eccentricities of planets d to h vs the normalised semi-major axis ratio $(a_{i+1}/a_i)\times[(j_i+1)/j_i]^{-2/3}$ of their resonances. As was described in the introduction, tidal evolution will drive the planets away from resonance so that the motion in this figure is towards the bottom-right corner: the eccentricities decrease and the separation increases. The big dots show the current configuration for planets e to h. It is clear that despite the age of the system the outermost pairs are still very close to resonance, with the normalised semi-major axis ratio between planets g and h being less than 1.009, while it is less than 1.005 for the other pairs. In our simulations the system likely started at higher eccentricity and a lower semi-major axis ratio and then was allowed to tidally evolve towards (and past) the current configuration.

\begin{figure*}
\resizebox{\hsize}{!}{\includegraphics[angle=0]{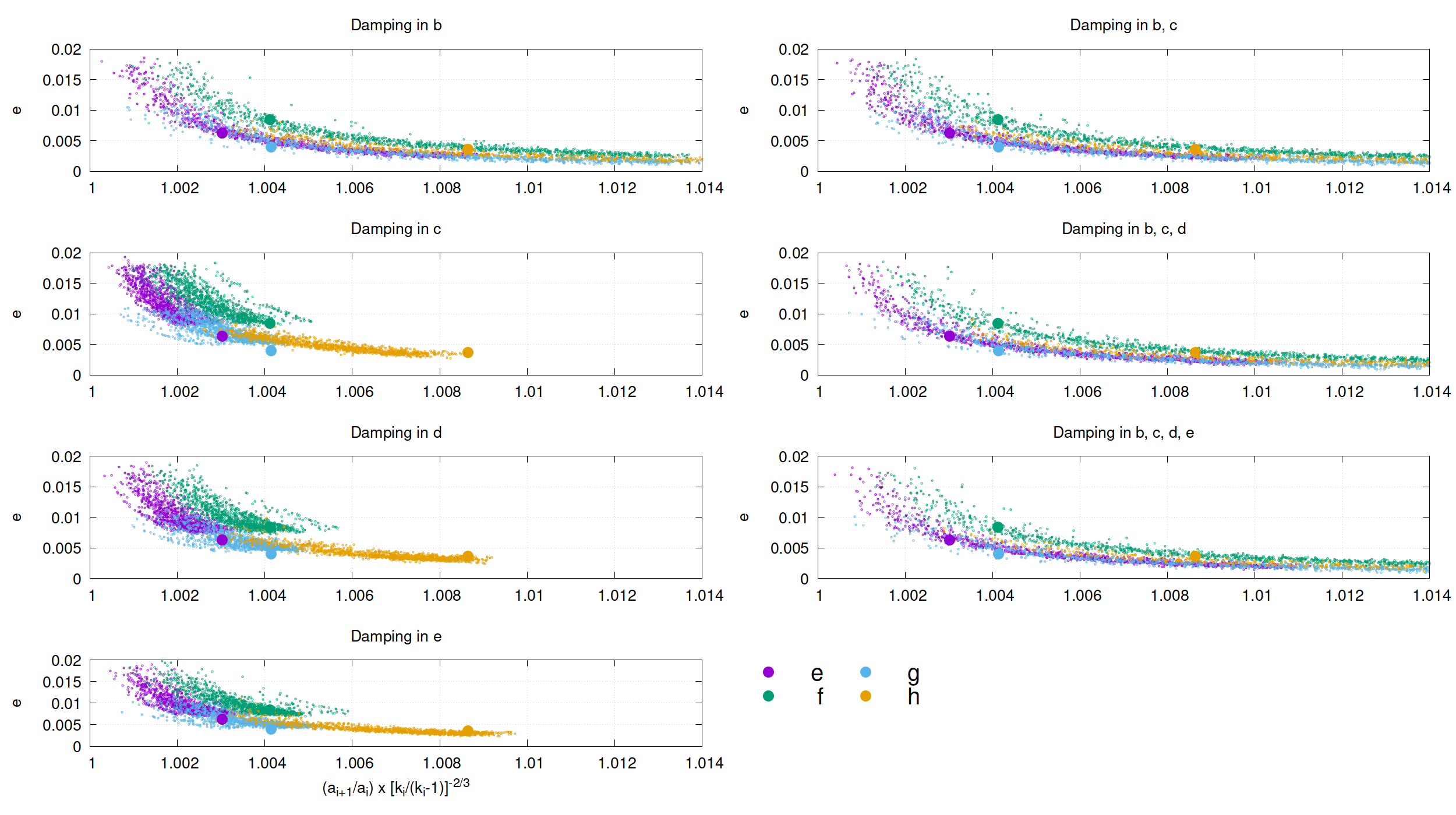}}
\caption{Tidal evolution on equilibrium curves for the current system with a five-planet resonant chain. The colours correspond to planets e--h as indicated in the legend. Large dots indicate the present configuration. The motion of the system is towards the right i.e. towards greater separation and lower eccentricities.}
\label{fig:aecc4}
\end{figure*}

\begin{figure}
\resizebox{\hsize}{!}{\includegraphics[angle=0]{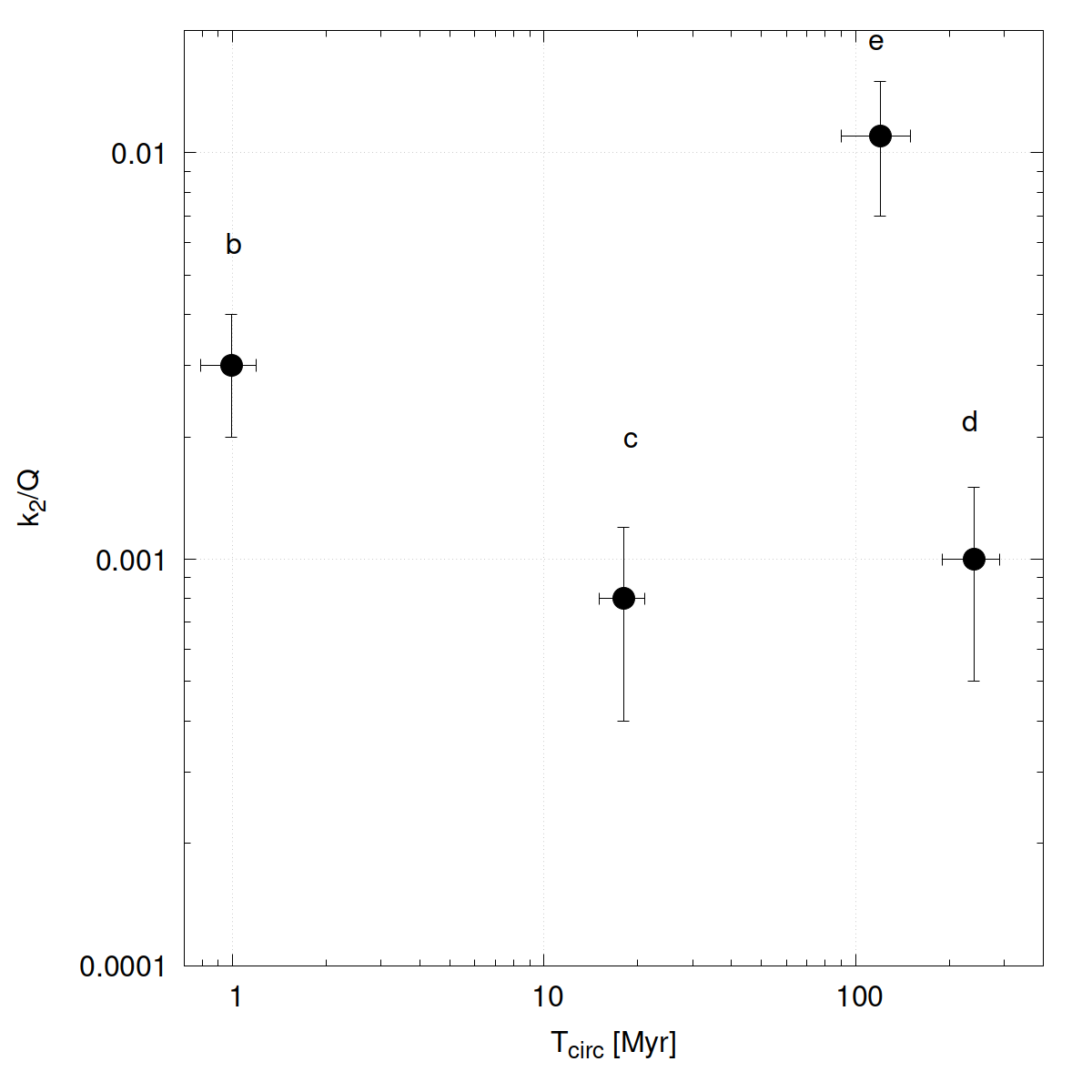}}
\caption{Long-term averaged tidal parameters for plants b to e obtained from dynamical analysis. The uncertainties are due to the different outcomes of the tidal simulations: different simulations result in different tidal circularisation timescales depending on which planet the damping occurs in.}
\label{fig:k2qdyn}
\end{figure}

\begin{figure}
\resizebox{\hsize}{!}{\includegraphics[angle=0]{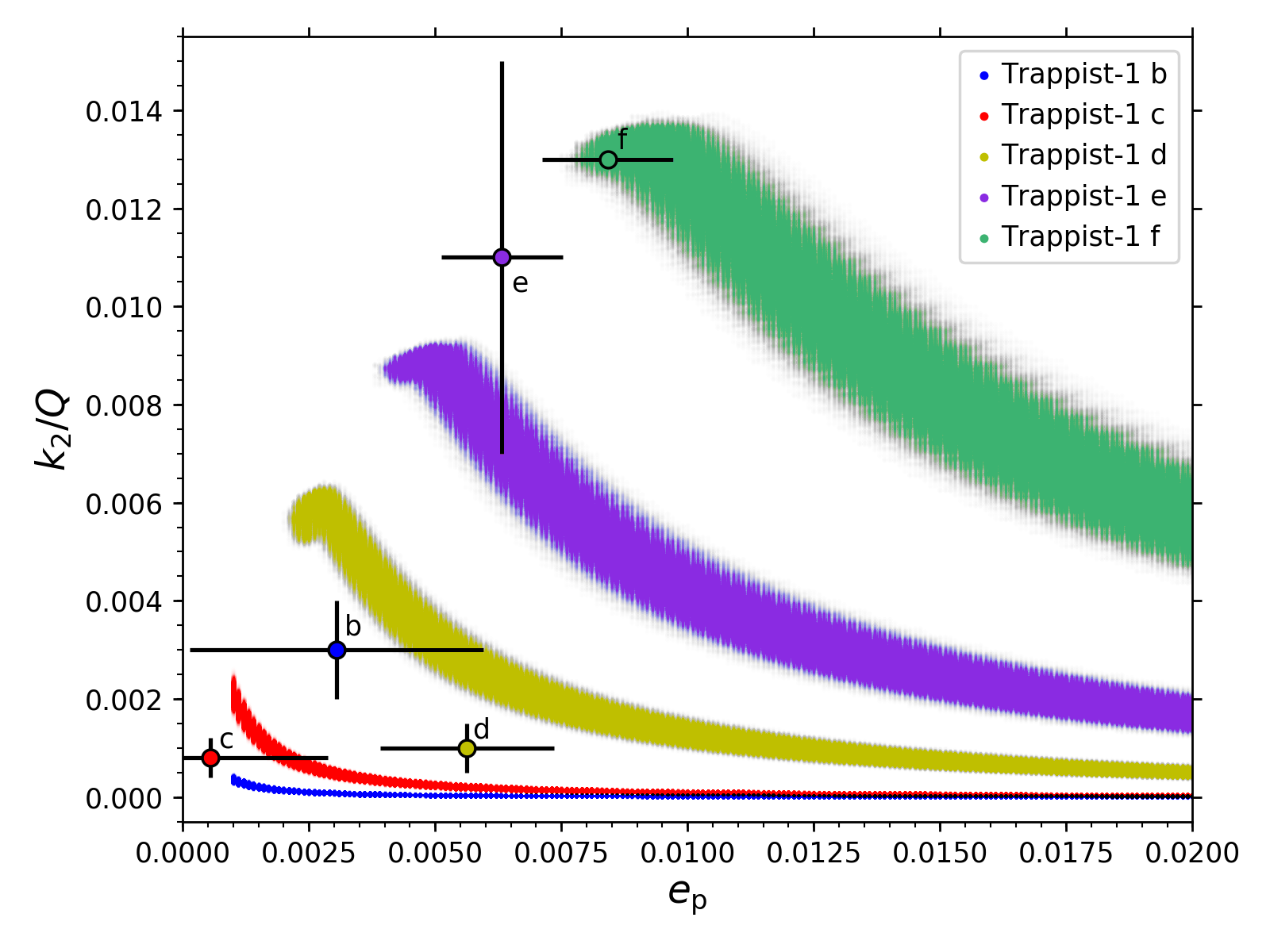}}
\caption{Tidal parameters for plants b to f obtained from interior modelling as a function of eccentricity (coloured diagonal bands). The results from the dynamical analyses are overplotted as small circles with error bars. The colour of the circle matches the colour of the band for each planet for easy comparison}. Horizontal error bars: range of eccentricity values based on the work of \citet{Agol2021}. Vertical error bars: possible range of the $k_2/Q$ tidal parameter based on our dynamical simulations (see also Fig.~\ref{fig:k2qdyn}).
\label{fig:k2qint}
\end{figure}

\begin{figure*}
\resizebox{\hsize}{!}{\includegraphics[angle=0]{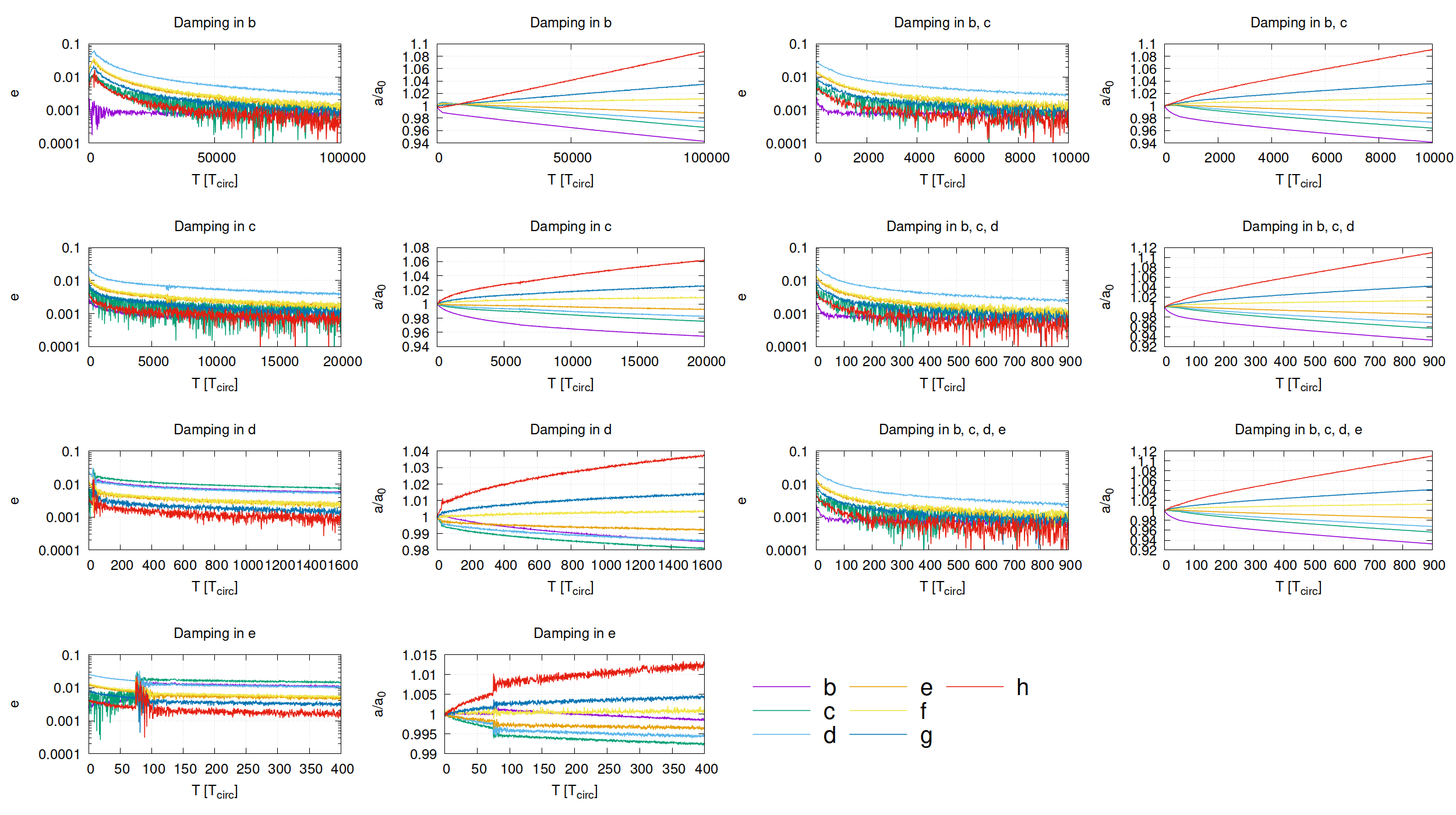}}
\caption{Tidal evolution starting from a (3:2)$^4$--4:3--3:2 Trappist-1 chain. Different panels show the resulting evolution of the eccentricities and the relative change in semi-major axis for different choices of tidal damping.}
\label{fig:ecc2}
\end{figure*}

\begin{figure*}
\resizebox{\hsize}{!}{\includegraphics[angle=0]{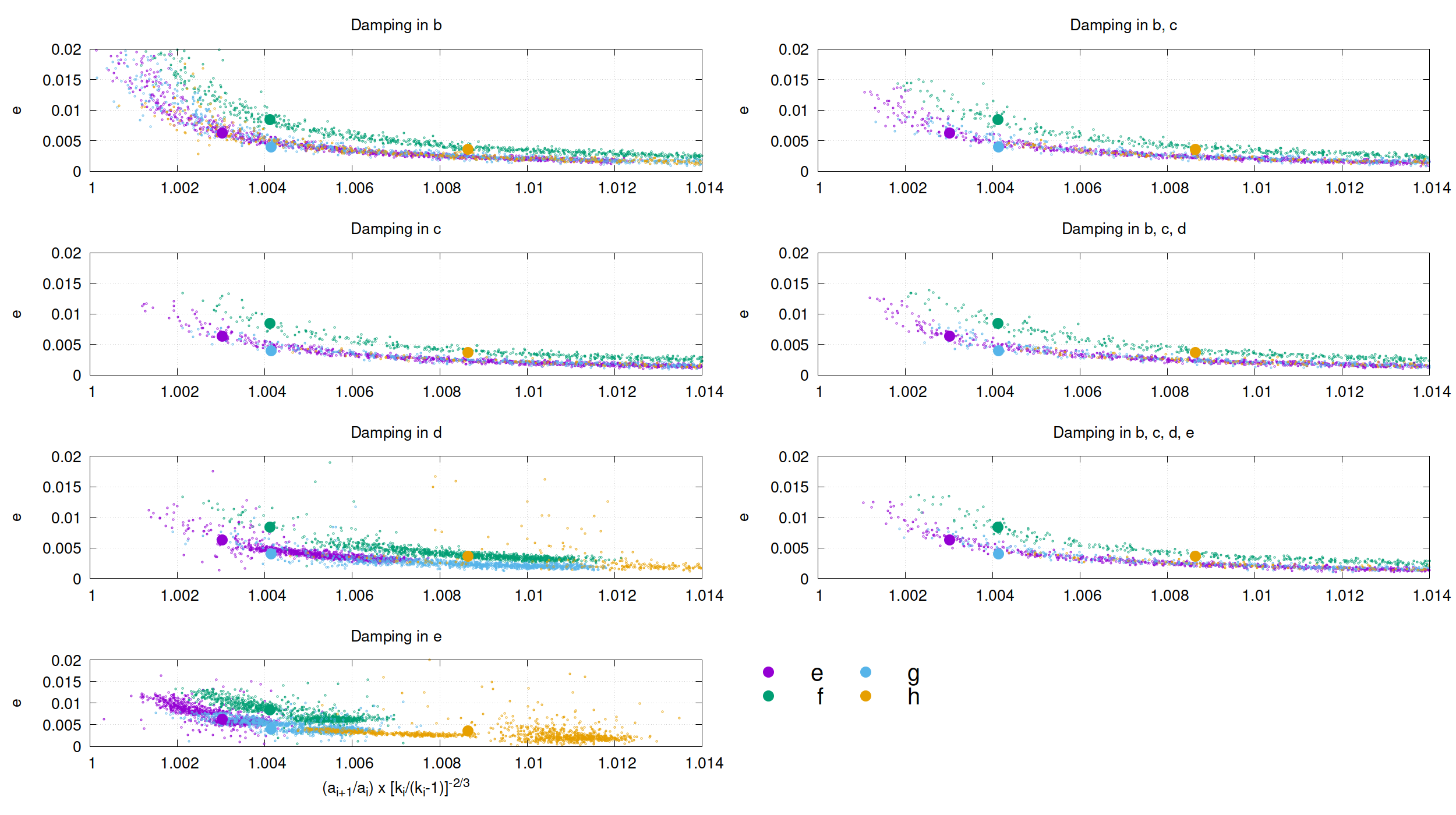}}
\caption{Tidal evolution on equilibrium curves for a (3:2)$^4$--4:3--3:2 chain with a five-planet resonant chain. The colours are indicated by the legend. Large dots indicate the present configuration. The motion of the system is towards the right i.e. towards greater separation and lower eccentricities.}
\label{fig:aecc2}
\end{figure*}

One question arises from these results: can we compute the long-term average tidal dissipation parameters of the planets from this evolution and compare them with those obtained from reasonable interior models as in \citet{Barr2018}?

Yes, it turns out that we can use the tidal evolution to compute the long-term $k_2/Q$ for the planets assuming that the outer five migrated to their current configuration starting from an initial configuration closer to resonance (Fig.~\ref{fig:aecc4}). Their mutual separation cannot be overshot and thus we can compute the tidal parameters from the time it takes to reach the current configuration of the outer five planets starting from the resonant chain after migration. We proceed as follows.

We compute the time it takes in the simulation for the eccentricities of e, f, g and h to decrease by a factor of the Euler's number, $e=2.71828...$. We do this for each simulation wherein the damping only happens in a single planet. The tidal parameters input into the simulations yield a simulation tidal damping timescale. From the eccentricity evolution, depending on the specific simulation and in which planet the damping occurs, we obtain the true damping timescale in one of the planets b to e by multiplying the simulation running time with the simulation damping timescale. The long-term value of $k_2/Q$ is then obtained from \citep[e.g.][]{BatyginMorbidelli2013}

\begin{equation}
    \Bigl(\frac{k_2}{Q}\Bigr)= \frac{2}{21n\tau_e}\frac{m_p}{M_*}\Bigl(\frac{a}{R_p}\Bigr)^{5},
\end{equation}
where $\tau_e$ is the true eccentricity damping timescale and $R_p$ is the radius of the planet. The result is displayed in Fig.~\ref{fig:k2qdyn}, where we plot the damping timescale versus the long-term $k_2/Q$ value. The uncertainties in both quantities are due to variations in outcome between simulations.

We compare the $k_2/Q$ values obtained from dynamical simulations with those obtained from simulations from interior modelling as a function of eccentricity. The results of these simulations for planets b to f are shown in Fig.~\ref{fig:k2qint}; the interior modelling simulations do not yield $k_2/Q$ values for planets g and h because the tidal dissipation is too weak. For each orbital eccentricity value that we tested there is a range of possible $k_2/Q$ parameters as a consequence of different possible interior structures and uncertainties in the planets' mass, radius and semi-major axis. The results are plotted with semi-transparent colours, making the less likely solutions (at the border or the ranges) appear in a shade of grey. The farther a planet is from the host star, the higher orbital eccentricity it needs to induce tidal heating in the body that is higher than heating from the star or from radioactive decay; this explains why the curves start at higher eccentricity values from planets b to f. The $k_2/Q$ parameter monotonically decreases with orbital eccentricity because the heat flux is a constant value. Horizontal error bars show the possible eccentricity range of each planet based \citet{Agol2021}, while vertical error bars represent the results of the dynamic simulation in accordance with Fig~\ref{fig:k2qdyn}. Apart from planets b and d, within uncertainty the values of $k_2/Q$ from dynamics and interior modelling match. Considering also the current orbital eccentricities of the planets, we get the following results for the other three planets. For planet c the interior models yield $k_2/Q = (8 \pm 4) \times 10^{-4}$ with orbital eccentricity between 0.0013 and 0.0028; for planet e the two models (dynamical and interior) combined yield a value of $k_2/Q$ in the range (0.007, 0.0092), which is narrower than that obtained only from dynamics or only based on interior modelling; and for planet f the interior models yield $k_2/Q$ in the range (0.0114,  0.0138). Just like in \citet{Brasser2019} the tidal parameter values show some differences with the outcome from dynamics, but given the uncertainties and the number of parameters involved the near agreement is encouraging. Indeed, the agreement is better than in \citet{Brasser2019}, most likely due to improved masses, radii and eccentricities of the planets, and a corresponding narrower range of interior compositions and amount of tidal heating.

\cite{Brasser2019} suggested from dynamical simulations that for planets b and c $k_2/Q\gtrsim 2\times10^{-4}$ and $k_2/Q \gtrsim 10^{-3}$ respectively, while interior models with planetary mass and radii data available at that time implied $k_2/Q$ values of a factor of a few lower. \cite{Bolmont2020} computed $k_2/Q$ for planet e for a variety of interior models and forcing frequencies, in which they have a homogenous interior and a layered one. At the current orbital frequency of planet e their computed value of $k_2/Q$ in the layered model is about a factor of five lower than ours derived from dynamics. For their homogenous model their derived value agrees with our interior modelling. For planet b, from their tidal evolution simulations \cite{HuangOrmel2022} suggest that $Q\approx 200k_2$, so that if $k_2/Q \approx 5\times 10^{-3}$, which is close to our value obtained from dynamics, but is much higher than those obtained from the interior models.

\subsection{Tidal evolution of a (3:2)$^4$ -- 4:3 -- 3:2 chain}
In this subsection we investigate whether can we reproduce the current configuration of the system from the more compact (3:2)$^4$--4:3--3:2 chain.

The tidal evolution of a (3:2)$^4$--4:3--3:2 chain is shown in Figs.~\ref{fig:ecc2} and Fig.~\ref{fig:aecc2}. In some cases -- such as when there is only damping in planet d, in planet e, or planets b and c -- the eccentricities sometimes increase before decreasing again. This is due to a pair of planets crossing a mean-motion resonance as they migrate divergently. We have found that in the cases where there is a lot of separation planets g and h cross their 5:3 resonance, while in the bottom panels (damping in planet e) planets b and c cross the 3:2 after about 70 circularisation times ($T_{\rm circ}$); the same happens when there is damping in planet d (third row, leftmost panels) after about 30 $T_{\rm circ}$. In contrast, in Fig.~\ref{fig:ecc4} we placed the planets closer to their current configuration and no resonances were crossed; the eccentricities of the five outer planets declined monotonically and the planets smoothly separate from each other. Fig.~\ref{fig:aecc2} is similar to Fig.~\ref{fig:aecc4}.

It is clear from Figs.~\ref{fig:ecc2} and~\ref{fig:aecc2} that dissipation in the planets alone does not reproduce the current configuration. Dissipation in the planets decreases the total energy but keeps the angular momentum constant. The orbital angular momentum is given by $L=\textstyle{\frac{m_p}{M_*}}\sqrt{a(1-e^2)}$. After the gas-driven migration into a (3:2)$^4$--4:3--3:2 chain the fraction of the total orbital angular momentum in the planets in our simulations is 15.2\%, 16.4\%, 5.2\%, 11.0\%, 18.7\%, 26.2\% and 7.3\% with the masses that we have employed. Thus, even though planet f is much farther away than planet c, it has a comparable amount of the total orbital angular momentum because the inner two planets are the most massive, and because of the weak $\sqrt{a}$ dependence on the distance (the eccentricity contribution is negligibly small). Furthermore, in the (3:2)$^4$--4:3--3:2 chain the inner four planets contain roughly half of the total orbital angular momentum. For the current configuration of the planets the fractions of the total orbital angular momentum are 14.6\%, 16.1\%, 5.3\%, 11.1\%, 19.0\%, 26.6\% and 7.3\% respectively for the set of representative initial conditions that we have chosen; thus the fraction of the inner two planets has decreased and that of the outer five has marginally increased. To reach their current orbits from a primordial (3:2)$^4$--4:3--3:2 chain and assuming that planet d experienced almost no migration the semi-major axis of planet b had to decrease by 10\% and that of planet c by 7\%. These reductions are minimum values: if planet d is allowed to migrate then the inner two planets would have initially resided even farther away from the star and their reduction in angular momentum would be even greater. From the bottom-right panel of Fig.~\ref{fig:ecc2} we see that having tidal dissipation in planets b to e causes a decrease in planet b's semi-major axis of 6\% to 0.94 and in planet c's by 4\% to 0.96 at the end of the simulation. This results in a corresponding increase in the semi-major axis of planet h of about 10\% and up to 4\% for planet g. In contrast, planets e and f generally do not migrate much (up to 2\% in extreme cases). Such large amounts of outward migration amongst the outer two planets occur to compensate for the inward migration of the inner planets, but it is inconsistent with the outer planets' current period ratio and resonant configuration (see Fig.~\ref{fig:aecc4}). The motion has to occur along the equilibrium curves and thus the large amount of inward migration of the innermost planets would not only require the outer five planets to have evolved to period ratios larger than observed, but also that their eccentricities are much lower than their current values. 

The current total orbital angular momentum is about 1.3\% lower than in the resonant (3:2)$^4$--4:3--3:2 chain after migration. Planet b's angular momentum now is 94.8\% of that in the (3:2)$^4$--4:3--3:2 chain chain after migration, while it is 96.5\% for planet c assuming that we keep the position of planet d fixed. This reduction in the orbital angular momentum of these two planets equals the 1.3\% reduction in the total angular momentum when using the original fractions because $0.948\times14.6\%+0.965\times16.5\%+5.3\%+11.1\%+19\%+26.6\%+7.3\% = 98.7\%$. Since angular momentum has to be strictly conserved for tidal dissipation in the planets we deem it unlikely, if not impossible, that tidal dissipation in the planets alone has resulted in the current configuration starting from a primordial (3:2)$^4$--4:3--3:2 chain. Testing simulations with dissipation in the outer three planets will not change that outcome because the planets evolve along the equilibrium curves. Furthermore, the long-term $k_2/Q$ values in the planets that we obtain from the tidal evolution with these initial conditions are the same as those presented in Fig.~\ref{fig:k2qdyn} within uncertainty.

The tidal parameters assumed here are may be extreme to keep simulation time short, but tests with tidal damping that is an order of magnitude weaker yields the same outcome due to the angular momentum conservation. 

\subsection{Inward migration of planets b and c}
One manner in which the inner two planets can lose angular momentum is through tidal dissipation in the star. For a constant $Q$ tidal model the semi-major axis evolution of the planets due to tidal dissipation in the star is given by \citep[e.g][]{murray99} 
\begin{equation}
    \frac{2}{13}a_f^{13/2}\Bigl[1-\Bigl(\frac{a_0}{a_f}\Bigr)^{13/2}\Bigr]=-3\Bigl(\frac{k_2}{Q}\Bigr)_*\Bigl(\frac{G}{M_*}\Bigr)^{1/2}m_pR_*^5t_*,
\end{equation}
where $t_*$ is the age of the star, $R_*$ is the stellar radius, $M_*$ is the stellar mass, $a_0$ is the initial semi-major axis and $a_f$ is the current value. We can solve this equation for $(k_2/Q)_*$. Starting planet b in a 3:2--3:2 with planets c and d, assuming that the age of the star $t_* = 7.6 \pm 2.2$~Gyr \citep{BM2017}, and that the stellar radius and rotation rate have stayed the same during this whole time yields $(k_2/Q)_* = 4.2^{+1.8}_{-1.2} \times 10^{-4}$; for planet c we get $(k_2/Q)_*=1.8^{+0.6}_{-0.4} \times 10^{-3}$, which is the same order of magnitude value. These values are much higher than that expected in solar-type stars, for which typically $(k_2/Q)_*<10^{-7}$ \citep[e.g][]{Penev2012,Ogilvie2014}. Even for lower-mass stars these high values of $(k_2/Q)_*$ are not expected, even though the tidal dissipation parameter decreases before the star reaches the main sequence \citep{Barker2020}. The uncertainties are mostly due to the uncertain age of the star.

\begin{figure*}
\resizebox{\hsize}{!}{\includegraphics[angle=0]{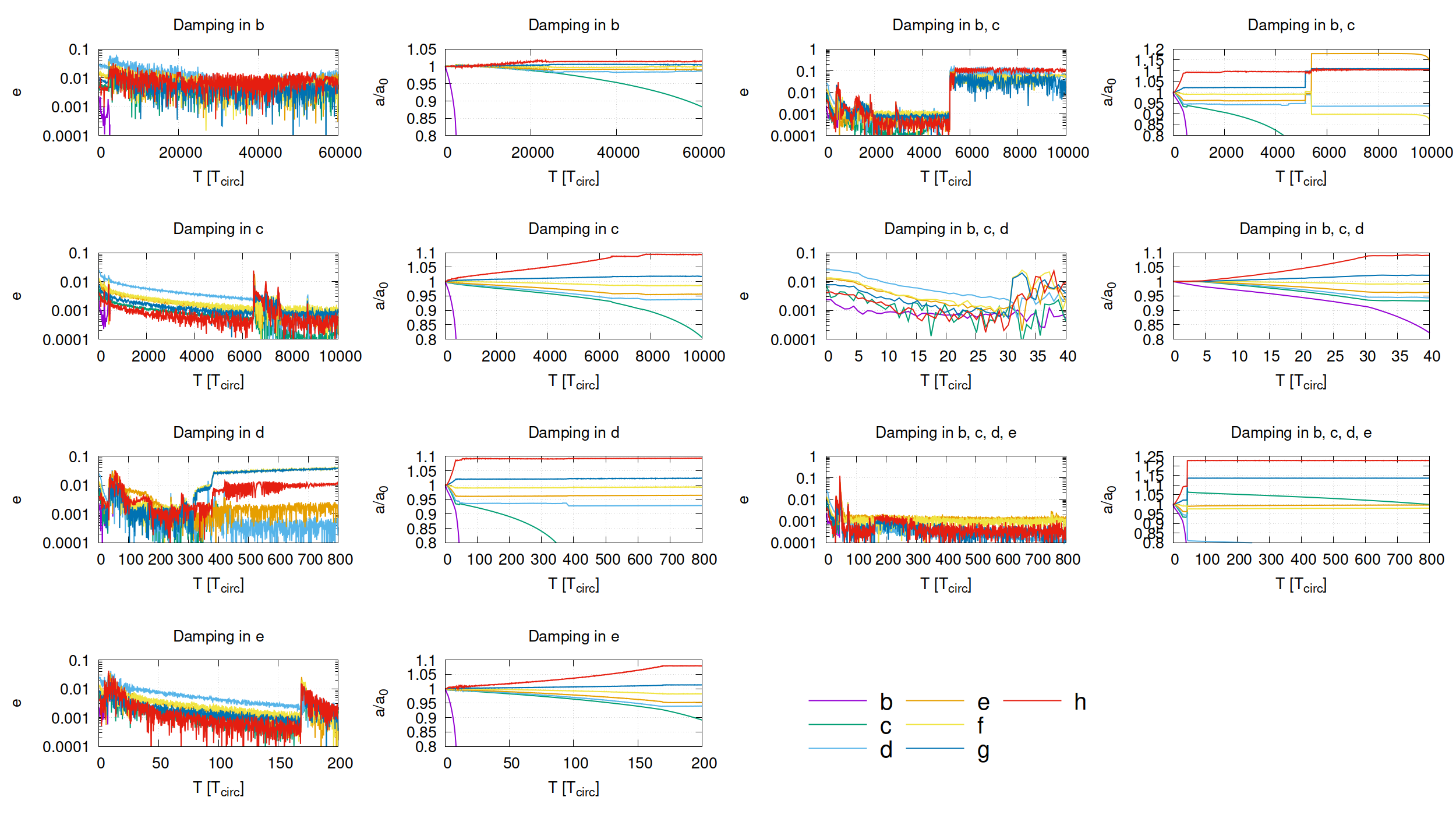}}
\caption{Tidal evolution starting with a fully resonant (3:2)$^4$--4:3--3:2 chain and with stellar tides included. The colours are indicated by the legend. Planet b rapidly detaches from the system and spirals towards the star, while eventually the same happens for planet c. The inward motion of planet c drags planet d and e with it, which causes planets f, g and h to separate from the rest of the system.}
\label{fig:aecc2s}
\end{figure*}

However, when we add dissipation in the star into our numerical simulations starting with a (3:2)$^4$--4:3--3:2 chain we run into a few problems. An example of the evolution is displayed in Fig.~\ref{fig:aecc2s}. First, the resonance between planets c and d drags planet d with planet c as it spirals towards the star. Eventually the pair will decouple and planet d will recede from the star because it is outside the corotation radius. However, by this time planet b has already spiralled too far inwards to be consistent with the current configuration. Second, by the time the c-d pair decouples and planet c begins to spiral towards the star, the g-h pair has crossed the 5:3 mean motion resonance due to the near-conservation of the angular momentum due to the resonant chain and that the planets try to evolve along the equilibrium curves.

The reason that planet b decouples quickly from the system and spirals inwards is because we were unable to migrate it deep into a resonance with planet c.

We when taking into account the reduction in stellar radius with time as predicted from isochrones by \cite{Baraffe2015} we still obtain a high long-term average value of $(k_2/Q)_*$ compared to that of solar-type stars.

\section{Discussion}
In the previous sections we argued that it is difficult to trap the planets into their current resonant configuration during an episode of gas-driven migration. Instead, the expected outcome is that the planets were caught in a chain of first-order mean-motion resonances \citep[e.g.][]{Ormel2017,HuangOrmel2022,Tey2022}. From such a configuration, such as the a (3:2)$^4$--4:3--3:2 chain, reaching the current configuration with dissipation in the planets seems unlikely. Yet there are a few potential caveats. For example, changing the dissipation parameters, or the strength of the dissipation in the planets relative to each other will not alter the outcome, because tidal dissipation in the planets conserves angular momentum, and the current configuration has less angular momentum than the (3:2)$^4$--4:3--3:2 chain.

Does the tidal model matter? Probably, but the differences between several models are not so great \citep[e.g.][]{}. At the moment we only possess one N-body code with a built-in tidal model, and thus we cannot test other models. However, the planets evolve along the equilibrium curves and the planetary tides conserve angular momentum. As such, we argue that changing the tidal model will not change the outcome because all tidal models conserve angular momentum; the only thing that could change is the evolution timescale for a specific value of the tidal parameters.

Do the initial conditions matter? Yes, but starting from a different resonant configuration means a different value of the total orbital angular momentum, which stays constant during the tidal evolution in the planets alone. Therefore, the precise initial conditions are unimportant unless they are very close to the current configuration. 

\section{Conclusions}
The TRAPPIST-1 system is lodged in a 5 or possibly even 7-planet resonant chain \citep{Luger2017,Agol2021}. As a consequence of this configuration, the tidal evolution of each planet cannot be calculated individually; rather the system evolves as one unit. Due to the resonant chain the tidal evolution proceeds along a set of equilibrium curves in semi-major axis--eccentricity space wherein the planets separate from each other as their forced eccentricities are damped by dissipation in the planets keeping the total angular momentum conserved. We show that the current 8:5--5:3--(3:2)$^2$--4:3--3:2 configuration cannot be reproduced from a primordial (3:2)$^4$--4:3--3:2 chain, so that an extra mechanism is required to decouple planets b and c from such a configuration. Dissipation in the star is one such mechanism, but with the tidal model employed here the decoupling of planet c from planet d takes longer than the time it takes for planet b to reach the star. As such it seems highly unlikely if not impossible to reproduce the current configuration from a (3:2)$^4$--4:3--3:2 chain using tides alone.

We also show that constraining the $k_2/Q$ tidal parameter is possible for some of the planets using both dynamical simulations on the orbits of the planets and interior structure models. For planet c we get $k_2/Q = (8 \pm 4) \times 10^{-4}$ with orbital eccentricity between 0.0013 and 0.0028. For planet e we get a range for $k_2/Q$ between 0.007 and 0.0092, and for planet f between 0.0114 and 0.0138.

\section*{Acknowledgements}
The authors thank Simon Grimm for sharing with us 10\,000 initial conditions of the system that are consistent with the TTV observations, and Inga Kamp and Floris van der Tak for useful discussion. AB acknowledges support from NASA Habitable Worlds grant 80NSSC20K1555. VD has been supported by the Hungarian National Research, Development, and Innovation Office (NKFIH) grant K-131508; the COFUND project oLife has received funding from the European Union's Horizon 2020 research and innovation programme under grant agreement No 847675.

%%%%%%%%%%%%%%%%%%%%%%%%%%%%%%%%%%%%%%%%%%%%%%%%%%
\section*{Data Availability}
Some of the code that we use is proprietary and we have been allowed to use it and modify it with permission. As such, the simulation codes and output can be obtained upon request.



\bibliographystyle{mnras}
\bibliography{example} % if your bibtex file is called example.bib

\begin{thebibliography}{}
\makeatletter
\relax
\def\mn@urlcharsother{\let\do\@makeother \do\$\do\&\do\#\do\^\do\_\do\%\do\~}
\def\mn@doi{\begingroup\mn@urlcharsother \@ifnextchar [ {\mn@doi@}
  {\mn@doi@[]}}
\def\mn@doi@[#1]#2{\def\@tempa{#1}\ifx\@tempa\@empty \href
  {http://dx.doi.org/#2} {doi:#2}\else \href {http://dx.doi.org/#2} {#1}\fi
  \endgroup}
\def\mn@eprint#1#2{\mn@eprint@#1:#2::\@nil}
\def\mn@eprint@arXiv#1{\href {http://arxiv.org/abs/#1} {{\tt arXiv:#1}}}
\def\mn@eprint@dblp#1{\href {http://dblp.uni-trier.de/rec/bibtex/#1.xml}
  {dblp:#1}}
\def\mn@eprint@#1:#2:#3:#4\@nil{\def\@tempa {#1}\def\@tempb {#2}\def\@tempc
  {#3}\ifx \@tempc \@empty \let \@tempc \@tempb \let \@tempb \@tempa \fi \ifx
  \@tempb \@empty \def\@tempb {arXiv}\fi \@ifundefined
  {mn@eprint@\@tempb}{\@tempb:\@tempc}{\expandafter \expandafter \csname
  mn@eprint@\@tempb\endcsname \expandafter{\@tempc}}}

\bibitem[\protect\citeauthoryear{{Agol} et~al.,}{{Agol}
  et~al.}{2021}]{Agol2021}
{Agol} E.,  et~al., 2021, \mn@doi [Planetary Science Journal]
  {10.3847/PSJ/abd022}, \href
  {https://ui.adsabs.harvard.edu/abs/2021PSJ.....2....1A} {2, 1}

\bibitem[\protect\citeauthoryear{{Baraffe}, {Homeier}, {Allard}  \&
  {Chabrier}}{{Baraffe} et~al.}{2015}]{Baraffe2015}
{Baraffe} I.,  {Homeier} D.,  {Allard} F.,   {Chabrier} G.,  2015, \mn@doi
  [\aap] {10.1051/0004-6361/201425481}, 577, A42

\bibitem[\protect\citeauthoryear{{Barker}}{{Barker}}{2020}]{Barker2020}
{Barker} A.~J.,  2020, \mn@doi [\mnras] {10.1093/mnras/staa2405}, 498, 2270

\bibitem[\protect\citeauthoryear{{Barr}, {Dobos}  \& {Kiss}}{{Barr}
  et~al.}{2018}]{Barr2018}
{Barr} A.~C.,  {Dobos} V.,   {Kiss} L.~L.,  2018, \mn@doi [\aap]
  {10.1051/0004-6361/201731992}, \href
  {https://ui.adsabs.harvard.edu/abs/2018A&A...613A..37B} {613, A37}

\bibitem[\protect\citeauthoryear{{Batygin} \& {Morbidelli}}{{Batygin} \&
  {Morbidelli}}{2013a}]{BatyginMorbidelli2013}
{Batygin} K.,  {Morbidelli} A.,  2013a, \mn@doi [\aj]
  {10.1088/0004-6256/145/1/1}, \href
  {https://ui.adsabs.harvard.edu/abs/2013AJ....145....1B} {145, 1}

\bibitem[\protect\citeauthoryear{{Batygin} \& {Morbidelli}}{{Batygin} \&
  {Morbidelli}}{2013b}]{2013A&A...556A..28B}
{Batygin} K.,  {Morbidelli} A.,  2013b, \mn@doi [\aap]
  {10.1051/0004-6361/201220907}, 556, A28

\bibitem[\protect\citeauthoryear{{Bolmont}, {Raymond}, {Leconte}, {Hersant}  \&
  {Correia}}{{Bolmont} et~al.}{2015}]{Bolmont2015}
{Bolmont} E.,  {Raymond} S.~N.,  {Leconte} J.,  {Hersant} F.,   {Correia} A.
  C.~M.,  2015, \mn@doi [\aap] {10.1051/0004-6361/201525909}, 583, A116

\bibitem[\protect\citeauthoryear{{Bolmont}, {Breton}, {Tobie}, {Dumoulin},
  {Mathis}  \& {Grasset}}{{Bolmont} et~al.}{2020}]{Bolmont2020}
{Bolmont} E.,  {Breton} S.~N.,  {Tobie} G.,  {Dumoulin} C.,  {Mathis} S.,
  {Grasset} O.,  2020, \mn@doi [\aap] {10.1051/0004-6361/202038204}, \href
  {https://ui.adsabs.harvard.edu/abs/2020A&A...644A.165B} {644, A165}

\bibitem[\protect\citeauthoryear{{Brasser}, {Barr}  \& {Dobos}}{{Brasser}
  et~al.}{2019}]{Brasser2019}
{Brasser} R.,  {Barr} A.~C.,   {Dobos} V.,  2019, \mn@doi [\mnras]
  {10.1093/mnras/stz1231}, \href
  {https://ui.adsabs.harvard.edu/abs/2019MNRAS.487...34B} {487, 34}

\bibitem[\protect\citeauthoryear{{Burgasser} \& {Mamajek}}{{Burgasser} \&
  {Mamajek}}{2017}]{BM2017}
{Burgasser} A.~J.,  {Mamajek} E.~E.,  2017, \mn@doi [\apj]
  {10.3847/1538-4357/aa7fea}, \href
  {https://ui.adsabs.harvard.edu/abs/2017ApJ...845..110B} {845, 110}

\bibitem[\protect\citeauthoryear{{Chambers}}{{Chambers}}{1999}]{Chambers1999}
{Chambers} J.~E.,  1999, \mn@doi [\mnras] {10.1046/j.1365-8711.1999.02379.x},
  304, 793

\bibitem[\protect\citeauthoryear{{Cresswell} \& {Nelson}}{{Cresswell} \&
  {Nelson}}{2008}]{2008A&A...482..677C}
{Cresswell} P.,  {Nelson} R.~P.,  2008, \mn@doi [\aap]
  {10.1051/0004-6361:20079178}, 482, 677

\bibitem[\protect\citeauthoryear{{Delisle}, {Laskar}, {Correia}  \&
  {Bou{\'e}}}{{Delisle} et~al.}{2012}]{Delisle2012}
{Delisle} J.~B.,  {Laskar} J.,  {Correia} A.~C.~M.,   {Bou{\'e}} G.,  2012,
  \mn@doi [\aap] {10.1051/0004-6361/201220001}, 546, A71

\bibitem[\protect\citeauthoryear{{Dobos}, {Barr}  \& {Kiss}}{{Dobos}
  et~al.}{2019}]{Dobos2019}
{Dobos} V.,  {Barr} A.~C.,   {Kiss} L.~L.,  2019, \mn@doi [\aap]
  {10.1051/0004-6361/201834254}, \href
  {https://ui.adsabs.harvard.edu/abs/2019A&A...624A...2D} {624, A2}

\bibitem[\protect\citeauthoryear{{Duncan}, {Levison}  \& {Lee}}{{Duncan}
  et~al.}{1998}]{Duncan1998}
{Duncan} M.~J.,  {Levison} H.~F.,   {Lee} M.~H.,  1998, \mn@doi [\aj]
  {10.1086/300541}, 116, 2067

\bibitem[\protect\citeauthoryear{Elkins-Tanton \& Seager}{Elkins-Tanton \&
  Seager}{2008}]{elkins08}
Elkins-Tanton L.~T.,  Seager S.,  2008, The Astrophysical Journal, 688, 628

\bibitem[\protect\citeauthoryear{{Gillon} et~al.,}{{Gillon}
  et~al.}{2017}]{Gillon2017}
{Gillon} M.,  et~al., 2017, \mn@doi [\nat] {10.1038/nature21360}, \href
  {https://ui.adsabs.harvard.edu/abs/2017Natur.542..456G} {542, 456}

\bibitem[\protect\citeauthoryear{{Goldreich} \& {Schlichting}}{{Goldreich} \&
  {Schlichting}}{2014}]{Goldreich2014}
{Goldreich} P.,  {Schlichting} H.~E.,  2014, \mn@doi [\aj]
  {10.1088/0004-6256/147/2/32}, 147, 32

\bibitem[\protect\citeauthoryear{{Grimm} et~al.,}{{Grimm}
  et~al.}{2018}]{Grimm2018}
{Grimm} S.~L.,  et~al., 2018, \mn@doi [\aap] {10.1051/0004-6361/201732233},
  \href {https://ui.adsabs.harvard.edu/abs/2018A&A...613A..68G} {613, A68}

\bibitem[\protect\citeauthoryear{{Henning}, {O'Connell}  \&
  {Sasselov}}{{Henning} et~al.}{2009}]{henning09}
{Henning} W.~G.,  {O'Connell} R.~J.,   {Sasselov} D.~D.,  2009, \mn@doi [ApJ]
  {10.1088/0004-637X/707/2/1000}, \href
  {http://adsabs.harvard.edu/abs/2009ApJ...707.1000H} {707, 1000}

\bibitem[\protect\citeauthoryear{{Henrard}}{{Henrard}}{1982}]{Henrard1982}
{Henrard} J.,  1982, \mn@doi [Celestial Mechanics] {10.1007/BF01228946}, \href
  {https://ui.adsabs.harvard.edu/abs/1982CeMec..27....3H} {27, 3}

\bibitem[\protect\citeauthoryear{{Huang} \& {Ormel}}{{Huang} \&
  {Ormel}}{2022}]{HuangOrmel2022}
{Huang} S.,  {Ormel} C.~W.,  2022, \mn@doi [\mnras] {10.1093/mnras/stac288},
  511, 3814

\bibitem[\protect\citeauthoryear{{Laskar}}{{Laskar}}{1993}]{Laskar1993}
{Laskar} J.,  1993, \mn@doi [Celestial Mechanics and Dynamical Astronomy]
  {10.1007/BF00699731}, 56, 191

\bibitem[\protect\citeauthoryear{{Levison} \& {Duncan}}{{Levison} \&
  {Duncan}}{1994}]{LD1994}
{Levison} H.~F.,  {Duncan} M.~J.,  1994, \mn@doi [\icarus]
  {10.1006/icar.1994.1039}, 108, 18

\bibitem[\protect\citeauthoryear{{Luger} et~al.,}{{Luger}
  et~al.}{2017}]{Luger2017}
{Luger} R.,  et~al., 2017, \mn@doi [Nature Astronomy]
  {10.1038/s41550-017-0129}, \href
  {https://ui.adsabs.harvard.edu/abs/2017NatAs...1E.129L} {1, 0129}

\bibitem[\protect\citeauthoryear{{Masset}, {Morbidelli}, {Crida}  \&
  {Ferreira}}{{Masset} et~al.}{2006}]{2006ApJ...642..478M}
{Masset} F.~S.,  {Morbidelli} A.,  {Crida} A.,   {Ferreira} J.,  2006, \mn@doi
  [\apj] {10.1086/500967}, \href
  {https://ui.adsabs.harvard.edu/abs/2006ApJ...642..478M} {642, 478}

\bibitem[\protect\citeauthoryear{{Murray} \& {Dermott}}{{Murray} \&
  {Dermott}}{1999}]{murray99}
{Murray} C.~D.,  {Dermott} S.~F.,  1999, {Solar System Dynamics}.
Cambridge University Press

\bibitem[\protect\citeauthoryear{{Ogihara}, {Duncan}  \& {Ida}}{{Ogihara}
  et~al.}{2010}]{Ogihara2010}
{Ogihara} M.,  {Duncan} M.~J.,   {Ida} S.,  2010, \mn@doi [\apj]
  {10.1088/0004-637X/721/2/1184}, \href
  {https://ui.adsabs.harvard.edu/abs/2010ApJ...721.1184O} {721, 1184}

\bibitem[\protect\citeauthoryear{{Ogihara}, {Kokubo}, {Nakano}  \&
  {Suzuki}}{{Ogihara} et~al.}{2022}]{Ogihara2022}
{Ogihara} M.,  {Kokubo} E.,  {Nakano} R.,   {Suzuki} T.~K.,  2022, \mn@doi
  [\aap] {10.1051/0004-6361/202142354}, \href
  {https://ui.adsabs.harvard.edu/abs/2022A&A...658A.184O} {658, A184}

\bibitem[\protect\citeauthoryear{{Ogilvie}}{{Ogilvie}}{2014}]{Ogilvie2014}
{Ogilvie} G.~I.,  2014, \mn@doi [\araa] {10.1146/annurev-astro-081913-035941},
  52, 171

\bibitem[\protect\citeauthoryear{{Ormel}, {Liu}  \& {Schoonenberg}}{{Ormel}
  et~al.}{2017}]{Ormel2017}
{Ormel} C.~W.,  {Liu} B.,   {Schoonenberg} D.,  2017, \mn@doi [\aap]
  {10.1051/0004-6361/201730826}, \href
  {https://ui.adsabs.harvard.edu/abs/2017A&A...604A...1O} {604, A1}

\bibitem[\protect\citeauthoryear{{Papaloizou}, {Szuszkiewicz}  \&
  {Terquem}}{{Papaloizou} et~al.}{2018}]{Papaloizou2018}
{Papaloizou} J.~C.~B.,  {Szuszkiewicz} E.,   {Terquem} C.,  2018, \mn@doi
  [\mnras] {10.1093/mnras/stx2980}, \href
  {https://ui.adsabs.harvard.edu/abs/2018MNRAS.476.5032P} {476, 5032}

\bibitem[\protect\citeauthoryear{{Penev}, {Jackson}, {Spada}  \&
  {Thom}}{{Penev} et~al.}{2012}]{Penev2012}
{Penev} K.,  {Jackson} B.,  {Spada} F.,   {Thom} N.,  2012, \mn@doi [\apj]
  {10.1088/0004-637X/751/2/96}, 751, 96

\bibitem[\protect\citeauthoryear{{Petrovich}, {Malhotra}  \&
  {Tremaine}}{{Petrovich} et~al.}{2013}]{Petrovich2013}
{Petrovich} C.,  {Malhotra} R.,   {Tremaine} S.,  2013, \apj, 770, 24

\bibitem[\protect\citeauthoryear{{Pichierri}, {Morbidelli}  \&
  {Crida}}{{Pichierri} et~al.}{2018}]{2018CeMDA.130...54P}
{Pichierri} G.,  {Morbidelli} A.,   {Crida} A.,  2018, \mn@doi [Celestial
  Mechanics and Dynamical Astronomy] {10.1007/s10569-018-9848-2}, 130, 54

\bibitem[\protect\citeauthoryear{{Pichierri}, {Batygin}  \&
  {Morbidelli}}{{Pichierri} et~al.}{2019}]{2019A&A...625A...7P}
{Pichierri} G.,  {Batygin} K.,   {Morbidelli} A.,  2019, \mn@doi [\aap]
  {10.1051/0004-6361/201935259}, 625, A7

\bibitem[\protect\citeauthoryear{{Puranam} \& {Batygin}}{{Puranam} \&
  {Batygin}}{2018}]{PB2018}
{Puranam} A.,  {Batygin} K.,  2018, \mn@doi [\aj] {10.3847/1538-3881/aab09f},
  155, 157

\bibitem[\protect\citeauthoryear{{Schoonenberg}, {Liu}, {Ormel}  \&
  {Dorn}}{{Schoonenberg} et~al.}{2019}]{Schoonenberg2019}
{Schoonenberg} D.,  {Liu} B.,  {Ormel} C.~W.,   {Dorn} C.,  2019, \mn@doi
  [\aap] {10.1051/0004-6361/201935607}, \href
  {https://ui.adsabs.harvard.edu/abs/2019A&A...627A.149S} {627, A149}

\bibitem[\protect\citeauthoryear{{Tamayo}, {Rein}, {Petrovich}  \&
  {Murray}}{{Tamayo} et~al.}{2017}]{Tamayo2017}
{Tamayo} D.,  {Rein} H.,  {Petrovich} C.,   {Murray} N.,  2017, \mn@doi [\apjl]
  {10.3847/2041-8213/aa70ea}, 840, L19

\bibitem[\protect\citeauthoryear{{Tanaka} \& {Ward}}{{Tanaka} \&
  {Ward}}{2004}]{Tanaka2004}
{Tanaka} H.,  {Ward} W.~R.,  2004, \mn@doi [\apj] {10.1086/380992}, \href
  {https://ui.adsabs.harvard.edu/abs/2004ApJ...602..388T} {602, 388}

\bibitem[\protect\citeauthoryear{{Teyssandier}, {Libert}  \&
  {Agol}}{{Teyssandier} et~al.}{2022}]{Tey2022}
{Teyssandier} J.,  {Libert} A.~S.,   {Agol} E.,  2022, \mn@doi [\aap]
  {10.1051/0004-6361/202142377}, 658, A170

\bibitem[\protect\citeauthoryear{Turbet, Bolmont, Ehrenreich, Gratier, Leconte,
  Selsis, Hara  \& Lovis}{Turbet et~al.}{2020}]{turbet2020}
Turbet M.,  Bolmont E.,  Ehrenreich D.,  Gratier P.,  Leconte J.,  Selsis F.,
  Hara N.,   Lovis C.,  2020, Astronomy \& Astrophysics, 638, A41

\bibitem[\protect\citeauthoryear{{Unterborn}, {Desch}, {Hinkel}  \&
  {Lorenzo}}{{Unterborn} et~al.}{2018}]{Unterborn2018}
{Unterborn} C.~T.,  {Desch} S.~J.,  {Hinkel} N.~R.,   {Lorenzo} A.,  2018,
  \mn@doi [Nature Astronomy] {10.1038/s41550-018-0411-6}, \href
  {https://ui.adsabs.harvard.edu/abs/2018NatAs...2..297U} {2, 297}

\bibitem[\protect\citeauthoryear{{Wisdom} \& {Holman}}{{Wisdom} \&
  {Holman}}{1991}]{WH1991}
{Wisdom} J.,  {Holman} M.,  1991, \mn@doi [\aj] {10.1086/115978}, 102, 1528

\bibitem[\protect\citeauthoryear{{{\v{S}}idlichovsk{\'y}} \&
  {Nesvorn{\'y}}}{{{\v{S}}idlichovsk{\'y}} \& {Nesvorn{\'y}}}{1996}]{SN1996}
{{\v{S}}idlichovsk{\'y}} M.,  {Nesvorn{\'y}} D.,  1996, \mn@doi [Celestial
  Mechanics and Dynamical Astronomy] {10.1007/BF00048443}, 65, 137

\makeatother
\end{thebibliography}

%\begin{thebibliography}{99}
%\end{thebibliography}

%%%%%%%%%%%%%%%%%%%%%%%%%%%%%%%%%%%%%%%%%%%%%%%%%%

%%%%%%%%%%%%%%%%% APPENDICES %%%%%%%%%%%%%%%%%%%%%

%\appendix

%\section{Tidal evolution starting with near-current initial conditions}


%%%%%%%%%%%%%%%%%%%%%%%%%%%%%%%%%%%%%%%%%%%%%%%%%%

% Don't change these lines
\bsp	% typesetting comment
\label{lastpage}
\end{document}